\documentclass[fleqn,usenatbib]{mnras}

\usepackage[T1]{fontenc}
\usepackage{ae,aecompl}

\usepackage{graphicx}   % Including figure files
\usepackage{amsmath}    % Advanced maths commands
\usepackage{amssymb}    % Extra maths symbols

\usepackage[usenames,dvipsnames]{xcolor}

\usepackage{amstext}
\usepackage{mathtools}
\usepackage{siunitx}
\usepackage{physics}
\usepackage{xspace}
\usepackage{cleveref}
\usepackage{pdflscape}
\usepackage{multirow}

\usepackage{newtxtext,newtxmath}
\DeclareSIUnit{\parsec}{pc}
\DeclareSIUnit{\pc}{pc}
\DeclareSIUnit{\year}{yr}
\newcommand{\Msun}{\ensuremath{\mathrm{M}_{\odot}}\xspace}

\newcommand{\limepy}{\textsc{limepy}\xspace}
\newcommand{\Nbody}{\(N\)-body\xspace}

\newcommand*\NGC[1]{NGC\thinspace{#1}}
\newcommand*\Messier[1]{M\thinspace{#1}}
\newcommand{\omegacen}{\(\omega\)\thinspace{Cen}\xspace}

\newcommand{\ra}{\ensuremath{r_{\mathrm{a}}}}

\newcommand{\rh}{\ensuremath{r_{\mathrm{h}}}}

\newcommand{\fbh}{\ensuremath{f_{\mathrm{BH}}}\xspace}
\newcommand{\Nrelax}{\ensuremath{N_{\mathrm{relax}}}\xspace}

\newcommand{\Gaia}{\textit{Gaia}\xspace}
\newcommand{\HST}{\textit{HST}\xspace}

\defcitealias{Dickson2023}{Paper~I}
\newcommand{\paperI}{\citetalias{Dickson2023}\xspace}

\title[Multimass modelling GCs - II. Black Holes]{
    Multimass modelling of Milky Way globular clusters
    - II. present-day black hole populations
}

\author[N. Dickson et al.]{
N. Dickson$^{1}$\thanks{E-mail: nolan.dickson@smu.ca}, P.J. Smith$^{1}$,
V. Hénault-Brunet$^{1}$, M. Gieles$^{2,3}$, H. Baumgardt$^{4}$
\\
$^{1}$Department of Astronomy and Physics, Saint Mary's University,
923 Robie Street, Halifax, NS B3H 3C3, Canada \\
$^{2}$ICREA, Pg. Llu\'{i}s Companys 23, E08010 Barcelona, Spain\\
$^{3}$Institut de Ci\`{e}ncies del Cosmos (ICCUB), Universitat de Barcelona (IEEC-UB), Mart\'{i} Franqu\`{e}s 1, E08028 Barcelona, Spain\\
$^{4}$School of Mathematics and Physics, The University of Queensland, St Lucia, QLD 4072, Australia
}

\date{Accepted XXX. Received YYY; in original form ZZZ}

\pubyear{2024}

\begin{document}
\label{firstpage}
\pagerange{\pageref{firstpage}--\pageref{lastpage}}
\maketitle

\begin{abstract}

Populations of stellar-mass black holes (BHs) in globular clusters (GCs)
influence their dynamical evolution and have important implications
on one of the main formation channels for gravitational wave sources.
Inferring the size of these populations remains difficult, however.
In this work, multimass models of 34 Milky Way GCs, first presented in
Dickson et al., are used to explore the present-day BH populations.
Direct constraints on both the total and visible mass
components provided by several observables allow these models to accurately
determine the distribution of the dark mass (including BHs) within clusters, as
we demonstrate in a proof-of-concept fitting of the models to mock
observations extracted from Monte Carlo cluster models.
New constraints on the BH population retained to the present-day in each
cluster are inferred from our models. We find that BH mass fractions ranging
from 0 to 1 per cent of the total mass are typically required to explain the
observations, except for \omegacen, for which we infer a mass fraction above
5 per cent, in agreement with previous works.
Relationships between the dark remnant populations and other cluster
parameters are examined, demonstrating a clear anti-correlation between the
amount of BHs and mass segregation between visible stars, as well as a
correlation between remnant mass fractions and the dynamical age of clusters.
Our inferred BH populations are in good agreement overall with other recent
studies using different methodologies, but with notable discrepancies for
individual clusters.

\end{abstract}


%---------------------------------------------------------------------------
% BODY OF PAPER
%---------------------------------------------------------------------------

%---------------------------------------------------------------------------
% Introduction
%---------------------------------------------------------------------------

% !TEX root = ./paper.tex

%---------------------------------------------------------------------------
\section{Introduction}\label{sec:introduction}
%---------------------------------------------------------------------------

    It has been suggested that dynamical formation of close stellar-mass black
    hole (BH) binaries in the dense cores of globular clusters (GCs) could be
    one of the main formation channels for gravitational wave sources
    from BH-BH mergers \citep[e.g.][]{PortegiesZwart2000,Rodriguez2016,
    Abbott2016a,Antonini2020b,Antonini2023}.
    The amount of dynamically formed BH-BH binaries in GCs and subsequent
    mergers, however, depends on many uncertain physical ingredients, such as
    the initial mass and number distribution of BHs (which itself is dependent
    on the stellar initial mass function and stellar evolution;
    \citealp[e.g.][]{Spera2015}) and
    the magnitude of natal kicks that BHs receive at the time of their
    formation in a supernova explosion, which can eject them from the cluster
    \citep{Chatterjee2017}, as well as
    the fraction of the initially retained BHs ejected due to
    dynamical encounters throughout a cluster’s life, which depends on its
    initial properties and dynamical evolution
    \citep[e.g.][]{Breen2013a,Hurley2016,Gieles2021}.

    The presence of massive BHs in GCs will have a large impact on their
    dynamical evolution and present-day structure.
    As the most massive objects in the system, any BHs which have not been
    ejected from the cluster by natal kicks will rapidly segregate to the
    cluster centre, forming a concentrated population of BHs in the core.
    While it was long argued that this dense subsystem would dynamically
    decouple from the rest of the GC \citep[e.g.][]{Spitzer1969,Sigurdsson1993},
    and in short order eject all of the BHs through strong dynamical
    interactions, recent work has shown that these BHs can be retained on much
    longer timescales \citep[e.g.][]{Breen2013a, Morscher2013, Morscher2015},
    and populations of BHs can be expected to survive in most clusters to the
    present day, depending on the initial cluster central densities.
    This theoretical work is complemented by recent detections of
    BH candidates in binaries within GCs, based on radio/X-ray emission
    (from accretion) or the radial velocity variations of a bright companion
    \citep{Strader2012,Millerjones2015,Giesers2018,Giesers2019}.
    Several studies examining detailed evolutionary dynamical models (such as
    \Nbody or Monte Carlo models) of clusters with an initial population of BHs
    have used these expected impacts on observable quantities, such as
    the absence of mass segregation among visible stars
    \citep[e.g.][]{Peuten2016,Alessandrini2016,Weatherford2018,Weatherford2020},
    an elevated central mass-to-light ratio \citep[e.g.][]{Baumgardt2019b} or a
    large effective radius of the cluster \citep{Torniamenti2023} and
    the presence of tidal tails \citep{Gieles2021}, to argue for
    the presence of retained populations of BHs in certain clusters at the
    present day.

    Alternatively, equilibrium modelling approaches, such as
    Jeans modelling \citep[e.g.][]{Kamann2014,Kamann2016,Abbate2019,
    Vitral2021,Vitral2022} and multimass distribution-function (DF) based models
    \citep[e.g.][]{Sollima2012, Sollima2016, Zocchi2019,Henault-Brunet2019,Henault-Brunet2020},
    have also been used to probe the dark remnant content of GCs (including
    BHs) through fitting models to observations of specific clusters
    while accounting for both visible and dark mass components.
    Rapid and flexible equilibrium models, such as DF models, allow for a more
    complete exploration of parameter space compared to more computationally
    expensive evolutionary models, and, through the application of statistical
    fitting techniques, the ability to very precisely reproduce the kinematics
    and structure of real, observed GCs.

    % ----------------------------------------------------------------------
    % Reference to paper I
    % ----------------------------------------------------------------------

    In \citet[][hereafter \paperI]{Dickson2023}, we presented
    multimass DF models fit to several observables (proper motions,
    line-of-sight velocities, stellar densities and mass functions), for
    34 Milky Way GCs.
    By inferring the global stellar mass functions of these clusters and
    simultaneously constraining their distributions of stellar remnants,
    the stellar initial mass function and its possible dependence on
    metallicity was examined, in particular in the high-mass
    regime (\(\geq 1\,\Msun\)) where stars in old GCs have
    evolved into stellar remnants by the present day.

    % ----------------------------------------------------------------------
    % This work
    % ----------------------------------------------------------------------

    In this work, these models are used to now
    examine in detail the remnant populations in this large sample of Milky Way
    GCs. In particular, the amount and distribution of BHs
    in each cluster are inferred from our models.
    The multimass \limepy models \citep{Gieles2015}, mass function evolution algorithm,
    observational datasets and model fitting procedures used are all
    restated briefly, from \paperI, in \Cref{sec:methods}.
    In \Cref{sec:validation}, we provide a proof of concept and show that our
    method is able to reliably recover the mass fraction in BHs in GCs by
    fitting our models to mock data extracted from snapshots of evolutionary
    models containing different amounts of BHs.
    The overall distributions of BHs in our sample of GCs
    are presented in \Cref{sec:black_hole_populations}, alongside an analysis of
    their relationships with other cluster parameters.
    In \Cref{sec:discussion} we discuss the implications of these results on
    the co-evolution of GCs and their BHs, and provide comparisons between our
    results and the inferred BH populations of other recent studies.
    Finally, we summarize our results and conclude in \Cref{sec:conclusions}.

%---------------------------------------------------------------------------
% Methods
%---------------------------------------------------------------------------

% !TEX root = ./paper.tex

%---------------------------------------------------------------------------
\section{Methods}\label{sec:methods}
%---------------------------------------------------------------------------

    In this section, the methodology used in \paperI to fit multimass dynamical
    models to a number of observables is restated briefly, with emphasis on the
    elements most crucial to inferring BH
    populations. For more detail on all procedures summarized here, we refer
    the reader to Sections 2 through 4 of \paperI.

\subsection{Models}

    To model the mass distribution of the globular clusters we use the \limepy
    multimass distribution-function (DF) based models \citep{Gieles2015}.
    DF based models are equilibrium models built around a distribution
    function which describes the phase-space particle density of
    stars and satisfies the collisionless Boltzmann equation.
    This DF is used to self-consistently solve for the system's potential
    (\(\phi(r)\)) using Poisson's equation, and to derive a variety of useful
    quantities for describing a GC, such as the projected
    velocity dispersion, the projected surface density and the total mass.

    \textit{Multimass} DF models, defined by a sum of component DFs for
    individual mass bins, allow for a more accurate description of real
    GCs, which are made up of a spectrum of stellar and remnant
    masses, and are necessary in order to account for the effects of mass
    segregation.

    Our multimass models are defined by 10 free parameters dictating the mass
    function and physical solution of the \limepy DF.
    The overall structure of these models is controlled by the (dimensionless)
    central potential parameter \(\hat{\phi}_0\), which defines how
    centrally concentrated the model is.
    To mimic the effects of the host galaxy's tides, the energy near the
    truncation radius is reduced, lowering the escape velocity of stars and
    making it easier to escape, with a sharpness of truncation defined by the
    parameter \(g\) (lower values resulting in a more abrupt truncation). 
    The models can be expressed in physical units by adopting relevant size and
    mass scales. In order to match observations, we opt to scale the models
    using the parameters for total cluster mass \(M\) and 3D
    half-mass radius \(\rh\).
    The \limepy models allow for velocity anisotropy through an extra angular
    momentum term in the DF, which produces an isotropic core followed by
    a degree of radial velocity anisotropy at a distance from the centre
    defined by the anisotropy radius parameter \(\ra\) and then returning to
    isotropy again near the truncation radius.
    The multimass version of the \limepy DF is defined in part by the
    mass-dependent velocity scale \(s_j\). This scaling captures the trend
    towards kinetic energy equipartition among stars of different masses and
    models the effects of mass segregation
    \citep{Gieles2015, Peuten2017, Henault-Brunet2019}, and is defined based
    on the parameter \(\delta\), such that \(s_j \propto m_j^{-\delta}\).

    The constituent discrete mass components which approximate the
    mass spectrum of a GC are represented in the multimass
    models by the total (\(M_j\)) and mean (\(m_j\)) masses of each
    mass bin.
    As DF-based models, such as \limepy, are equilibrium, instantaneous
    ``snapshot'' models, and do not directly simulate any temporal
    astrophysical processes during their computation, we must instead
    incorporate a separate prescription for stellar evolution from an initial
    mass function, over the age of the cluster, to the present-day stellar and
    remnant mass functions.
    In keeping with the formulation of canonical IMFs
    \citep[e.g.][]{Kroupa2001}, we use a 3-component broken power law, with
    power-law ``slopes'' of each component given by the free parameters
    \(\alpha_1,\,\alpha_2,\,\alpha_3\) (with break masses at 0.5 and 1 \Msun
    and bounded between 0.1 and 100 \Msun).
    To evolve the population of stars to the present day we follow the algorithm
    first described by \citet{Balbinot2018} and expanded upon in the
    \texttt{ssptools} library\footnote{Available at
    \url{https://github.com/SMU-clusters/ssptools}} and \paperI.
    The amount of stars which evolve off the main sequence over the lifetime of
    the clusters is dictated by a set of equations based on interpolated
    Dartmouth Stellar Evolution Program models \citep{Dotter2007,Dotter2008}.
    The types and masses of the stellar remnants formed by these evolved stars
    are then determined based on their initial mass, metallicity and
    initial-final mass relation (IFMR).
    The white dwarf (WD) IFMR, including the maximum initial mass which will
    form a WD, is interpolated from the MIST 2018 isochrones
    \citep{Dotter2016,Choi2016}.
    The BH IFMR, as well as the minimum initial
    mass required to form a BH, is interpolated directly from a grid of stellar 
    evolution library (SSE) models \citep{Banerjee2020}, using the rapid
    supernova scheme \citep{Fryer2012}.
    Stars with initial masses between the WD and BH precursor
    masses are assumed to form neutron stars (NS) with a mass of \(1.4\,\Msun\).

    The algorithm then must account for the loss of BHs through two different
    channels.
    Firstly, the ejection of, primarily low-mass, BHs through natal kicks is
    simulated. Beginning with the assumption that the kick velocity is drawn
    from a Maxwellian distribution with a dispersion of \(\SI{265}{\km\per\s}\)
    \citep{Hobbs2005} and scaled down by \(1-f_{\mathrm{fb}}\),
    where \(f_{\mathrm{fb}}\) is the fallback fraction, which we
    interpolated from the same grid of SSE models used for the BH IFMR. The
    fraction of BHs retained, in each mass bin, is then found by integrating
    the Maxwellian kick velocity distribution from 0 to the system initial
    escape velocity, which we compute as
    \(v_{\mathrm{esc}} = 2 \sqrt{2 \phi_0}\).
    BHs are also ejected over time from the core of GCs due to dynamical
    interactions with one another \citep[e.g.][]{Breen2013a,Breen2013b}. This
    is simulated through the direct removal of BHs, beginning with the heaviest
    mass bins (with larger gravitational interaction cross-sections)
    through to the lightest \citep[e.g.][]{Morscher2015,Antonini2020a}, until
    the combination of mass in BHs lost through both ejection channels leads to
    a final retained mass in BHs equal to the percentage of the
    initial mass in BHs for the given IMF specified by the BH mass retention
    fraction parameter (\(\mathrm{BH}_{\mathrm{ret}}\)).

    Finally, the heliocentric distance to the GCs \(d\) is introduced as a free
    parameter, to allow for the conversion between projected, linear model
    units and the angular units of observations.

\subsection{Fitting models to observations}

    In \paperI, best-fitting model parameters for 34 Milky Way GCs were
    determined through the comparison of the phase-space
    distribution of stars in the models to observations of
    GC structure and kinematics.
    These clusters were chosen primarily to address the main
    hypothesis of \paperI (i.e. the possible metallicity dependence
    of the IMF) and were selected based on the quantity
    and quality of kinematic and mass function data available),
    however they also allow us to
    conduct a census of BH populations in a significant sample of Milky
    Way GCs (although our sample may be biased against low-mass, low-density
    and metal-rich clusters, as discussed in \cref{sec:discussion}).

    A variety of observational datasets were used to fit all chosen GCs,
    providing direct constraints on the phase-space distribution of visible
    stars and the overall mass of the cluster, and in turn providing
    indirect constraints on the amount and distribution of dark mass (in both
    faint low-mass stars and dark remnants). This is possible due to the fact
    that the distribution of the
    different dark and visible mass components are arranged with limited
    flexibility and linked, thanks to partial energy equipartition.
    Though BHs may represent only up to a few per cent of the total mass of a
    GC, they can actually dominate the mass density in the central regions of
    a cluster, with significant dynamical effects, and thus can be probed.
    Details of the datasets used for each
    cluster can be found in Appendix A of \paperI, but we summarize the main
    sources below.

    Radial profiles of proper motion (PM) and line-of-sight (LOS) velocity
    dispersions of cluster stars are used to constrain the internal kinematics
    of each cluster and thus its total (visible and dark) mass.
    PMs in both the radial and tangential directions
    also provide constraints on the degree of velocity anisotropy in the
    cluster, which is important given the degeneracy between anisotropy and
    central dark mass \citep[e.g.][]{Zocchi2017}.
    PM dispersion (radial and tangential) profiles were computed in \paperI
    using \Gaia (DR3; \citealp{GaiaCollaboration2022}) proper motions for all
    clusters. This data was supplemented with PM dispersion profiles in the
    cores of most clusters from Hubble Space Telescope data
    \citep[\HST,][]{Watkins2015,Libralato2022}.
    LOS velocity dispersion profiles are taken from compilations of various
    ground-based \citep{Baumgardt2018,Kamann2018,Dalgleish2020} and
    \Gaia \citep{Baumgardt2019a} datasets.

    Radial profiles of the projected stellar number density in all GCs
    provide vital constraints on the spatial distribution and concentration of
    the clusters stars. The density profiles for all clusters are taken from
    \citet{deBoer2019}, consisting of combined {\it Gaia} star counts in the outskirts
    and \HST counts \citep{Miocchi2013} or ground-based surface-brightness
    profiles \citep[SBPs,][]{Trager1995} in the central regions.

    Finally, constraints on the global present-day mass function of
    the clusters, the degree of mass segregation and the total mass in visible
    stars are provided by \HST datasets, from which local stellar mass functions
    were extracted. The mass function data
    for each cluster is taken from \citet{Baumgardt2023}, consisting of a
    stellar counts, based on a large amount of archival \HST data, in radial
    annuli and mass bins.
    The compilation of photometry in each cluster is made up of several \HST
    fields at varying distances from the cluster centres, typically covering
    stars within a mass range of \(\sim\SI{0.16}-\SI{0.8}{\Msun}\).

% \subsection{Model Fitting}

    The models are constrained by these datasets in order to provide
    best-fitting values and posterior probability distributions of the
    model parameters that describe each
    cluster, determined through Bayesian parameter estimation
    techniques\footnote{Latest model fits, cumulative mass profiles and
    black hole distributions for all clusters are available online at
    \url{https://github.com/nmdickson/GCfit-results}}.
    The posterior probability distributions of all parameters are sampled
    using dynamic nested sampling, through the \texttt{dynesty} software
    package \citep{Speagle2020}.
    All fitting is carried out using the software library and fitting
    pipeline \texttt{GCfit}\footnote{Available at
    \url{https://github.com/nmdickson/GCfit}}.
    For a discussion of the model fits and parameter posterior distributions,
    see \paperI.
    It should be noted that the latest version of \texttt{GCfit}, which includes
    various small improvements, such as to the accuracy of the mass function
    likelihood, was used to refit all of the models presented in this paper,
    however the conclusions of \paperI remain unaffected.

    The posterior probability distributions of the model-derived quantities used
    throughout this work, such as the black hole mass fractions
    (\(\fbh=M_{\mathrm{BH}}/M_{\mathrm{cluster}}\)), are
    constructed based on the models representing the set of weighted
    posterior samples retrieved from the nested sampler.

% !TEX root = ./paper.tex

%---------------------------------------------------------------------------
\section{Validation of BH population inference}
\label{sec:validation}
%---------------------------------------------------------------------------

    In order to test the reliability of our method in inferring BH populations,
    we first apply it to simulated observations from Monte Carlo
    models with known BH populations.

    In order to explore a number of models with similar properties as real
    Milky Way clusters, we select snapshots from the existing grid of Cluster
    Monte Carlo \citep[CMC;][]{Rodriguez2022} models presented in
    \citet{Kremer2020b}\footnote{Available at
    \url{https://cmc.ciera.northwestern.edu/home}}. We select the snapshots
    using the same methodology as \citet{Rui2021a} which we briefly summarize
    here.

    The selections are based on the SBPs of \citet{Trager1995} and the
    velocity dispersion profiles (VDP) compiled by
    \citet{Baumgardt2017}, \citet{Baumgardt2018}
    and \citet{Baumgardt2023}\footnote{Available at
    {\url{https://people.smp.uq.edu.au/HolgerBaumgardt/globular}
    \label{foot:baum_cat}}}.
    We search for snapshots that are a good match to any clusters from the
    \citet[][2010 edition]{Harris} catalogue present in both the VDP and SBP
    compilations, leaving us with about 100 clusters to match to snapshots.
    We first use the metallicities from \citet[][2010 edition]{Harris} and the
    present-day galactocentric radii from \citet{Baumgardt2019a} to select the
    subset of models which are closest to the true values for each cluster.

    From this subset, we then
    search every model for snapshots that match suitably well to
    a given cluster's observed SBP and VDP
    simultaneously \footnote{See \citet{Rui2021a} for details on how the
    model profiles are extracted from the CMC snapshots
    for comparison with the observations.}.
    In order to select a snapshot, we adopt a threshold of
    \(s \equiv \max \left(\tilde{\chi}_{\mathrm{SBP}}^2,
    \tilde{\chi}_{\mathrm{VDP}}^2\right) < 10\)
    for the "fitting heuristic" \(s\) of \citet{Rui2021a}, which describes
    the goodness-of-fit of a snapshot based on the \(\tilde{\chi}^2\)
    statistic between the observations and the interpolated model profiles.
    We have found that a threshold of \(s < 10\) provides
    an acceptable balance between the number and fit quality of the retained
    snapshots.
    While a number of snapshots passing this filter have an apparently poor
    overall fit to one or both of the observational profiles, we opt to
    still include these snapshots in the sample, as
    our goal is not to select only snapshots which are perfect matches to
    specific real clusters but instead to build a sample of
    snapshots that are qualitatively
    similar to the Milky Way clusters examined in this work.
    For each cluster covered by the observational datasets we select the single
    best-fitting snapshot, where one exists.

\subsection{Mock observations}

    The search described above results in a sample of 53
    CMC model snapshots, representative of Milky Way GCs.
    From these we next extract synthetic observations
    designed to emulate the real observational data
    used to constrain the models examined in this work.

    We place each cluster at its respective heliocentric distance as reported
    by \citet{Baumgardt2021} and then use the \mbox{\textsc{cmctoolkit}} library
    \citep{Rui2021a, cmctoolkit} to calculate projected positions and
    velocities as well as simulated photometry for objects in each snapshot.

\subsubsection{Number density profiles}

    We extract projected number density profiles from the models,
    designed to emulate those of \citet{deBoer2019}.
    We select all stars brighter than \Gaia \(G = 20\), sort them
    into 50 radial bins, with equal numbers of
    stars, and  calculate the number density in each
    radial bin. All densities are assigned a Poisson counting error.
    The \citet{deBoer2019} profiles are combinations of \Gaia star counts
    in the outer regions and \HST and archival SBPs in the inner regions, where
    crowding becomes an issue, however we
    find that using the same \(G < 20\) cut over
    the entire radial extent of the cluster results in
    well-sampled profiles which cover a similar radial extent
    and have similar uncertainties to the \citet{deBoer2019} profiles.

\subsubsection{Proper motion dispersion profiles}

    We extract two sets of PM dispersion profiles for each snapshot,
    in order to represent the performance of the two different
    sources of PM observations used.

    In the inner regions, we seek to emulate the performance of the \HST based
    PM dispersion profiles of \citet{Libralato2022}.
    We select stars within the central \SI{100}{\arcsecond} of the cluster to
    mimic the footprint of an \HST ACS footprint, and limit our
    selection to stars within \(15 < V < 18\).
    We split the stars into radial bins containing at least 120
    stars each, up to a maximum of five bins. This provides sufficient
    radial coverage of the cluster while still allowing us to construct
    profiles for distant clusters, where limited numbers of stars pass the
    magnitude cut.
    We assume a typical uncertainty of \SI{0.1}{mas\per\year} on all stars.
    Within each bin we compute the mean velocity and velocity dispersion along
    with their associated uncertainties, assuming the velocities are drawn from
    a Gaussian error distribution, using MCMC. This is repeated for
    both the radial and tangential components of PM.
    The median and \(1 \sigma\) values of the dispersion in each bin
    are used going forward.

    In the outer regions, we seek to
    emulate the \Gaia DR3 based profiles of \citet{Vasiliev2021}.
    We base our magnitude cuts on their profiles, selecting all stars in the
    \(13 < G < 19\) range outside of the innermost
    \SI{100}{\arcsecond}, to avoid overlapping with the \HST profiles.
    We assign each star an uncertainty in proper motion based on
    its \(G\) band magnitude using the calibrations provided in Table 4 of
    \citet{Lindegren2021}, allowing us to replicate the performance of the
    \Gaia DR3 catalogue.
    We again bin the stars using the same conditions as in the inner \HST
    profiles, and calculate the velocity dispersion in each bin
    using the same method, again for both radial and tangential components.

\subsubsection{Line-of-sight velocity dispersion profiles}

    In addition to the PM dispersion profiles, we also
    extract LOS velocity dispersion profiles, designed to
    emulate those presented by \citet{Baumgardt2017}, \citet{Baumgardt2018} and
    \citet{Baumgardt2023}\(^{\ref{foot:baum_cat}}\).
    As the compilation of velocity dispersions that
    make up these profiles consist of several
    different inhomogeneous datasets, with varying precisions, we adopt the
    simplifying assumption of a typical uncertainty of
    \SI{1}{\kilo\metre\per\second} on all observed stars.
    We limit this dataset to only giants brighter than \(V=17\), which is
    typical of the datasets used in the observed compilations.
    We again sort the stars into
    several radial bins, requiring at least 70 stars per bin, up to a maximum
    of 10 bins, and compute the velocity dispersion
    for each in the same way as for the PM profiles.

\subsubsection{Stellar mass functions}

    We extract stellar mass function data for each of our snapshots, designed to
    emulate the datasets presented in \citet{Baumgardt2023}. These datasets
    consists of star counts, binned by stellar mass, extracted from archival
    \HST observations.
    While the real \HST fields are distributed somewhat
    randomly around each GC according to the various goals of each proposal for
    which they were originally observed,
    in general there is typically at least one
    exposure centred on the cluster centre and a number of fields placed
    outside of the central region.
    For simplicity, we opt to place one field over the centre of the cluster,
    covering a range of \SI{0}{\arcminute} - \SI{1.6}{\arcminute} in
    projected angular separation from the centre, and two outer
    annuli at radial distances of \SI{2.5}{\arcminute} and \SI{5}{\arcminute},
    each sized such that they cover the same area as the central field.
    The central field is split into 4 annuli, giving us a total of 6
    annular fields covering the central, intermediate and outer regions of the
    cluster.
    In many of the more well-studied clusters in our sample, such as
    \NGC104 and \omegacen, the large number of observed \HST fields actually
    provides much better coverage of the clusters than our fields simulated
    here, and thus the results of this section may actually be
    conservative.
    In each of these fields we extract stellar counts separately,
    in bins of stellar mass with widths of \SI{0.1}{\Msun}.

    In the real datasets, the faintest stars (lowest stellar masses) for
    which stellar counts can be extracted with reasonable completeness
    (\(> 90\) per cent) is a function of crowding.
    To replicate this effect in our synthetic mass functions we
    construct an empirical relation between the surface number density
    and the lowest observable mass within a field.
    We use \NGC104 as the basis for this calibration
    because it covers a wide range of number densities from its core to the
    outskirts and has a large number of \HST fields for which the mass function
    was extracted.
    We extract star counts in each field, down to a lower mass
    limit calculated by the above relation and up to the main sequence turn-off,
    replicating the performance of the observed \HST stellar counts.

    We assign Poisson counting errors to our stellar counts, though to reflect
    the scatter we typically see in the real data we also inflate these errors
    by a factor of \(F=3\) and re-sample each point within the errors,
    resulting in mass functions that are very similar to those of
    \citet{Baumgardt2023} (see Section 4.1.1.3 of \paperI for a description
    of this \(F\) parameter and its motivation).

\subsection{Validation results}

    After extracting the synthetic datasets, we then directly
    apply our model fitting method (as described in \Cref{sec:methods} and
    \paperI) and compare the resulting inferred BH mass fractions to
    the known BH population in the CMC models from which the mock
    data was extracted. 

\begin{figure*}
        \centering
        \includegraphics[width=\linewidth]{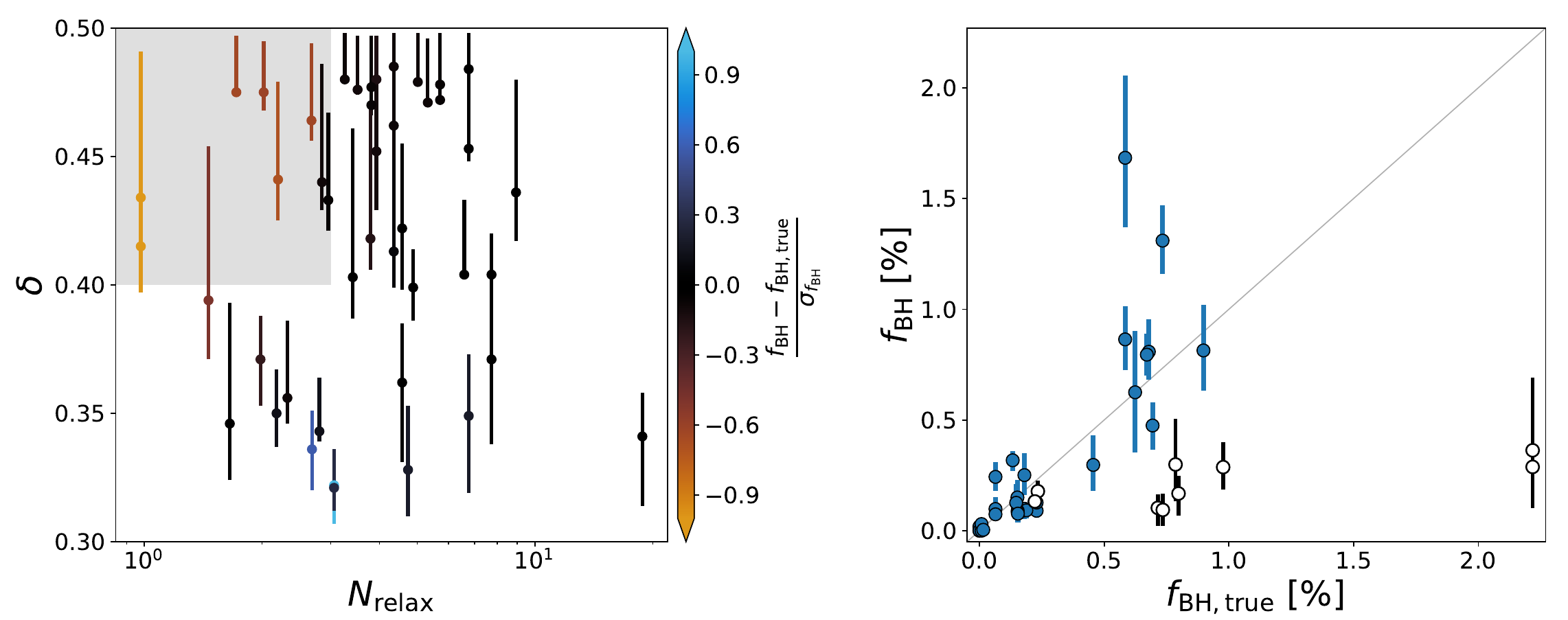}
        \caption{Left panel: Validation snapshots plotted in the
                 \(\delta - \Nrelax\) plane. A selection of dynamically young
                 clusters with high inferred values of \(\delta\) is shown by
                 the grey shaded region (\(\Nrelax<3\), \(\delta>0.4\)).
                 The points are coloured based on the number of
                 \(\sigma_{f_{\mathrm{BH}}}\) the
                 median inferred \fbh is away from the true value (where
                 \(\sigma_{f_{\mathrm{BH}}}\) is the width of the posterior for
                 \fbh). Under this colour scheme, snapshots for which we
                 underpredict (overpredict) \fbh are increasingly orange (blue).
                 Right panel:  The \fbh values inferred based on the multimass
                 model fits to mock observations extracted from CMC models,
                 against the true values in those validation snapshots
                 (\(f_{\mathrm{BH, true}}\)). Snapshots which fall within the
                 shaded region in the left panel are plotted with unfilled
                 markers in the right panel, demonstrating that they also
                 largely represent the snapshots for which we significantly
                 underpredict \fbh.
        \label{fig:delta-N}}
    \end{figure*}

\begin{figure}
        \centering
        \includegraphics[width=\linewidth]{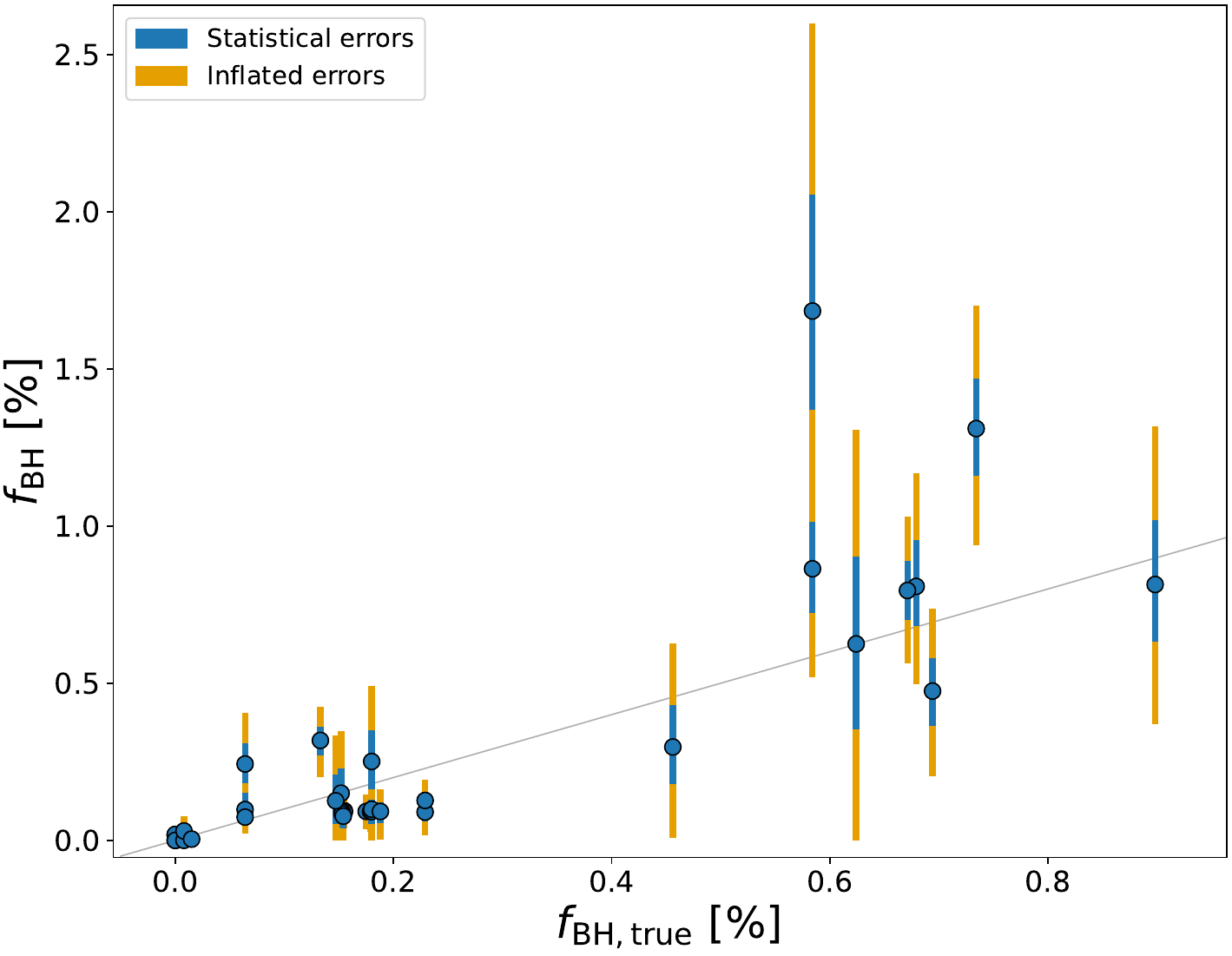}
        \caption{The \fbh values inferred based on the
                 mock observations extracted from CMC models, against the
                 true values in those models (\(f_{\mathrm{BH, true}}\)).
                 The one-to-one line is shown in grey, representing perfect
                 agreement.
                 The median and \(1 \sigma\) values, based solely on the
                 statistical uncertainties from the fit, are shown in blue.
                 The inflated errors, based on the nuisance parameter
                 \(\epsilon\) (\(\Delta = \epsilon\ \sigma_{\fbh}\)),
                 are shown in orange.}
        \label{fig:val-comp}
    \end{figure}

    As in \paperI, we discard any obviously poor fits. We also discard any
    snapshots which have much smaller datasets (mostly consisting of
    snapshots matched to very distant clusters).
    This leaves us with 44 final snapshots with datasets
    of similar quality to the Milky Way clusters we study in this work, and
    with model fits which satisfyingly reproduce the
    mock observations and recover the various cluster parameters well, such as
    the total mass, which we generally recover within \(\sim 10\) per cent
    and the half-mass radius, which we generally recover within
    uncertainties.
    Our inferred values of \fbh, compared to the true values for our
    collection of snapshots, are shown in \Cref{fig:delta-N} (right panel),
    \Cref{fig:val-comp}, and \Cref{table:val_results}.

    While in general our fits satisfyingly recover the
    mass fraction in BHs, we find a number of snapshots for which we
    underpredict the black hole mass fraction. A common feature of these
    problematic snapshots is their dynamical age. All of the mock clusters for
    which we significantly underpredict \fbh have an age that is less than
    3 times their present-day half-mass relaxation time.
    In other words, they are dynamically very young. There are also, however,
    many dynamically young clusters for which correctly recover \fbh.
    
    To understand this behaviour, we plot in the left panel of
    \Cref{fig:delta-N} all of the mock clusters used in the validation of our
    models on the \(\delta - \Nrelax\) plane (where \Nrelax is the ratio of
    cluster age to present-day half-mass relaxation time),
    coloured by the distance of their median inferred \fbh value from the true
    value. As \(\delta\) is a measure of the degree of energy equipartition in
    our multimass models (with typical values of \(\delta\approx0.5\)
    for mass-segregated, dynamically evolved clusters), we should normally
    expect dynamically young clusters (\(\lesssim 3\,\Nrelax\)) to have lower
    values of \(\delta\).
    The mock clusters for which we significantly underpredict \fbh are clearly
    concentrated in the upper-left corner of the \(\delta - \Nrelax\) plane,
    highlighted with the grey-shaded rectangle in the left panel of
    \Cref{fig:delta-N}. Their value of \(\delta\) is inconsistent with our
    expectations for these systems.

    Previous comparisons of the multimass \limepy models used here with
    \(N\)-body models \citep{Peuten2017} have confirmed this intuition that
    dynamically young clusters should have lower values of
    \(\delta\) (with \(\delta\sim0.35\)) and have also shown that models with
    any significant black hole population should have similarly low values of
    \(\delta\) until they eject most of their black holes (after
    several relaxation times).
    Therefore we should not expect to find any clusters in the upper-left or
    lower-right sections of the \(\delta - \Nrelax\) plane and any
    model fits which fall into these regions are worthy of some suspicion, even
    though it is not immediately clear why the data for some of these
    dynamically young clusters would prefer models with values of \(\delta\)
    close to 0.5.
    Simple test fits with \(\delta\) forced to a lower value
    (\(\delta \leq 0.35\)) have shown that a lower value of \(\delta\) can
    result in an increase in the inferred \fbh for these snapshots, as
    generally expected.

    Within our validation sample there are a handful of problematic
    clusters in the upper-left, suspect region of the \(\delta - \Nrelax\)
    plane. In the right panel of \Cref{fig:delta-N}, we show our inferred
    values of \fbh compared to the true values for our collection of snapshots,
    with the snapshots that fall into this problematic region plotted with
    unfilled markers.
    It can be clearly seen that, if these clusters are ignored, the correlation
    between the inferred and true black hole mass fraction is even stronger.
    In \Cref{fig:val-comp}, we plot the inferred values of \fbh compared to the
    true values but now ignoring these problematic snapshots and showing the
    \(1 \sigma\) statistical error bars (in blue) for the retained mock
    clusters.

    With these caveats in mind we can turn to the sample of real
    clusters we investigate in this work to look for clusters that fall into
    this suspect region. In the real sample only three clusters out of the
    34 fall into this region of relatively high \(\delta\) and low
    \(\Nrelax\) (compared to 9 of the 44 validation snapshots). The relative
    lack of problematic fits in the real sample compared to the mock sample
    highlights the fact that our mock sample is not a perfect imitation of the
    population of MW clusters we investigate in this work and is limited by the
    constraints of the CMC grid.
    One of these problematic clusters is \NGC3201, which is discussed further
    below (\Cref{subsubsec:NGC3201}). The other two are \NGC104 (47 Tuc) and
    \NGC5139 (\(\omega\) Cen), two of the most massive clusters in our sample,
    which still possess notable BH populations despite their high \(\delta\). 
    Our results for these two well-studied clusters concurs well with other
    literature values (see the discussion of each in
    \Cref{sec:clusters_of_interest}). One real cluster in our sample falls into
    the lower-right portion of \(\delta - \Nrelax\).
    This is \NGC6624, a core-collapsed cluster which is discussed further in
    \Cref{sec:core_collapsed_clusters}, and for which we argue that we
    incorrectly infer a non-zero number of black holes.

    As may be expected, the statistical uncertainties
    derived solely from the fitting procedure seem to underestimate
    the real uncertainties slightly.
    Our fitting procedure operates under the assumption that our models are a
    perfect representation of the data, and as such, in reality, may underestimate
    the true errors.
    It has been shown that multimass DF models, such as those used here, may
    underestimate the uncertainties when compared to more flexible models,
    such as Jeans models \citep{Henault-Brunet2019}, which could be indicative
    of systematic errors not captured in the statistical uncertainties and
    limitations in the ability of these models to perfectly reproduce the data.

    In an attempt to quantify this underestimation, we search for the factor
    by which the statistical uncertainties on our inferred values of
    \fbh need to be inflated to make them fully consistent
    with a one-to-one relation with the true model values.
    We define a nuisance parameter
    \(\epsilon\) by which we multiply the
    statistical errors \(\sigma_{\fbh}\) to determine the total,
    inflated error \(\Delta\) on each inferred \fbh.
    We then fit a fixed, one-to-one (i.e. slope of 1, intercept of 0) line
    through the points in \Cref{fig:val-comp}, assuming a Gaussian likelihood
    and allowing \(\epsilon\) to vary freely and inflate the uncertainties on
    the inferred \fbh.
    This fit results in a value of \(\epsilon=2.5\substack{+0.3 \\ -0.3}\).
    We also show the inflated \(1 \sigma\) error bars
    on the inferred \fbh in \Cref{fig:val-comp}
    (orange), demonstrating the additional uncertainty needed to make our
    inferred values consistent with the true values and illustrating
    the typical uncertainties expected when applying our method to real data.

    Overall, this comparison with mock observations extracted from dynamical
    simulations lends confidence in the ability of our methods to correctly
    recover the mass fraction in BHs in GCs, with the important note that the
    statistical uncertainties on these inferred values may
    underestimate the true uncertainties by up to 
    a factor of \(\sim2.5\), and with a reminder that
    dynamically young clusters with inferred values of \(\delta \sim0.5\)
    should be treated with caution.

%---------------------------------------------------------------------------
% Results
%---------------------------------------------------------------------------

% !TEX root = ./paper.tex

%---------------------------------------------------------------------------
\section{Black Hole Populations}\label{sec:black_hole_populations}
%---------------------------------------------------------------------------

    %-----------------------------------------------------------------------
    % Intro to Black Holes chapter
    %-----------------------------------------------------------------------

    We will now explore the populations of black holes (and other dark
    remnants) inferred from our best-fitting models.
    We will examine the distribution of the total mass, mass fraction and
    amount of black holes found from these fits,
    and explore possible correlations present between the remnant
    populations and other cluster properties. We will also discuss in more
    detail the case of `core-collapsed' clusters and how we model them.

    %-----------------------------------------------------------------------
    % BH mass and number distributions
    %-----------------------------------------------------------------------

    \begin{table*}
    \renewcommand*{\arraystretch}{1.4}
    \centering
    \footnotesize
    \begin{tabular}{llllll}
    \hline
    Cluster          & \(\fbh\ \left[\%\right]\)             & \({M}_{\mathrm{BH}}\ \left[M_\odot\right]\) & \({N}_{\mathrm{BH}}\)           & \(\bar{m}_{\mathrm{BH}}\ \left[\Msun\right]\) & \(f_{\mathrm{remn}}\ \left[\%\right]\) \\
    \hline
    NGC104          & \(0.046\substack{+0.017 \\ -0.009}\)  & \(420\substack{+150 \\ -80}\)               & \(60\substack{+20 \\ -10}\)     & \(6.7\)                                       & \(46.9\substack{+0.3 \\ -0.3}\) \\
    NGC288          & \(0.09\substack{+0.09 \\ -0.06}\)     & \(80\substack{+80 \\ -50}\)                 & \(6\substack{+5 \\ -4}\)        & \(14.1\)                                      & \(56\substack{+2 \\ -2}\) \\
    NGC362\(^\ast\) & \(0.070\substack{+0.010 \\ -0.017}\)  & \(190\substack{+30 \\ -50}\)                & \(14\substack{+2 \\ -3}\)       & \(14.0\)                                      & \(45.1\substack{+0.6 \\ -0.6}\) \\
    NGC1261          & \(0.14\substack{+0.07 \\ -0.07}\)     & \(300\substack{+100 \\ -100}\)              & \(19\substack{+10 \\ -9}\)      & \(14.0\)                                      & \(39\substack{+2 \\ -2}\) \\
    NGC1851          & \(0.060\substack{+0.011 \\ -0.006}\)  & \(200\substack{+40 \\ -20}\)                & \(15\substack{+3 \\ -2}\)       & \(13.2\)                                      & \(41.0\substack{+0.5 \\ -0.6}\) \\
    NGC2808          & \(0.08\substack{+0.01 \\ -0.02}\)     & \(700\substack{+100 \\ -200}\)              & \(47\substack{+10 \\ -10}\)     & \(15.7\)                                      & \(36.2\substack{+0.5 \\ -0.6}\) \\
    NGC3201          & \(0.008\substack{+0.026 \\ -0.008}\)  & \(10\substack{+50 \\ -10}\)                 & \(0.9\substack{+3.2 \\ -0.9}\)  & \(14.3\)                                      & \(46\substack{+1 \\ -1}\) \\
    NGC5024          & \(0.11\substack{+0.15 \\ -0.08}\)     & \(600\substack{+800 \\ -400}\)              & \(40\substack{+60 \\ -30}\)     & \(14.1\)                                      & \(43\substack{+1 \\ -2}\) \\
    NGC5139          & \(5.9\substack{+0.2 \\ -0.2}\)        & \(182000\substack{+5000 \\ -6000}\)         & \(12400\substack{+700 \\ -300}\) & \(14.8\)                                      & \(50.3\substack{+0.9 \\ -0.3}\) \\
    NGC5272          & \(0.3\substack{+0.1 \\ -0.1}\)        & \(1500\substack{+500 \\ -600}\)             & \(110\substack{+40 \\ -40}\)    & \(13.4\)                                      & \(46.9\substack{+0.8 \\ -0.7}\) \\
    NGC5904          & \(0.05\substack{+0.06 \\ -0.04}\)     & \(200\substack{+200 \\ -200}\)              & \(20\substack{+20 \\ -10}\)     & \(10.9\)                                      & \(55.7\substack{+1.0 \\ -0.9}\) \\
    NGC5986          & \(0.03\substack{+0.04 \\ -0.03}\)     & \(80\substack{+100 \\ -80}\)                & \(6\substack{+7 \\ -6}\)        & \(13.8\)                                      & \(53\substack{+1 \\ -1}\) \\
    NGC6093          & \(0.62\substack{+0.03 \\ -0.04}\)     & \(1800\substack{+100 \\ -100}\)             & \(129\substack{+9 \\ -9}\)      & \(14.1\)                                      & \(53.3\substack{+0.8 \\ -0.8}\) \\
    NGC6121          & \(0.09\substack{+0.05 \\ -0.03}\)     & \(80\substack{+50 \\ -30}\)                 & \(6\substack{+3 \\ -2}\)        & \(13.4\)                                      & \(64.6\substack{+0.9 \\ -0.8}\) \\
    NGC6171          & \(0.07\substack{+0.04 \\ -0.04}\)     & \(40\substack{+30 \\ -30}\)                 & \(3\substack{+2 \\ -2}\)        & \(13.9\)                                      & \(66.6\substack{+1.0 \\ -0.9}\) \\
    NGC6205          & \(0.4\substack{+0.2 \\ -0.2}\)        & \(1800\substack{+800 \\ -700}\)             & \(130\substack{+50 \\ -50}\)    & \(13.9\)                                      & \(55\substack{+1 \\ -1}\) \\
    NGC6218          & \(0.14\substack{+0.01 \\ -0.02}\)     & \(140\substack{+10 \\ -20}\)                & \(10.0\substack{+0.8 \\ -1.4}\) & \(14.1\)                                      & \(61.4\substack{+0.7 \\ -0.7}\) \\
    NGC6254          & \(0.09\substack{+0.08 \\ -0.07}\)     & \(200\substack{+200 \\ -100}\)              & \(14\substack{+11 \\ -9}\)      & \(13.2\)                                      & \(57.0\substack{+0.8 \\ -0.9}\) \\
    NGC6266\(^\ast\) & \(0.14\substack{+0.01 \\ -0.01}\)     & \(1060\substack{+80 \\ -70}\)               & \(130\substack{+10 \\ -10}\)    & \(8.2\)                                       & \(52\substack{+2 \\ -2}\) \\
    NGC6341          & \(0.2\substack{+0.1 \\ -0.1}\)        & \(600\substack{+400 \\ -300}\)              & \(50\substack{+30 \\ -20}\)     & \(12.4\)                                      & \(52\substack{+1 \\ -1}\) \\
    NGC6352          & \(0.062\substack{+0.008 \\ -0.010}\)  & \(57\substack{+6 \\ -8}\)                   & \(5.2\substack{+0.6 \\ -0.7}\)  & \(10.9\)                                      & \(59\substack{+2 \\ -2}\) \\
    NGC6362          & \(0.03\substack{+0.04 \\ -0.03}\)     & \(30\substack{+50 \\ -30}\)                 & \(2\substack{+3 \\ -2}\)        & \(12.4\)                                      & \(65.1\substack{+1.0 \\ -0.9}\) \\
    NGC6366          & \(0.07\substack{+0.08 \\ -0.05}\)     & \(20\substack{+30 \\ -20}\)                 & \(2\substack{+2 \\ -1}\)        & \(13.0\)                                      & \(55\substack{+2 \\ -2}\) \\
    NGC6397\(^\ast\) & \(0.019\substack{+0.001 \\ -0.001}\)  & \(21\substack{+2 \\ -2}\)                   & \(3.1\substack{+0.2 \\ -0.2}\)  & \(6.7\)                                       & \(57.2\substack{+0.8 \\ -0.8}\) \\
    NGC6541\(^\ast\) & \(0.054\substack{+0.008 \\ -0.011}\)  & \(130\substack{+20 \\ -30}\)                & \(12\substack{+2 \\ -2}\)       & \(10.1\)                                      & \(53.3\substack{+0.8 \\ -0.8}\) \\
    NGC6624\(^\ast\) & \(0.48\substack{+0.02 \\ -0.03}\)     & \(490\substack{+20 \\ -20}\)                & \(46\substack{+2 \\ -2}\)       & \(10.8\)                                      & \(70.6\substack{+0.7 \\ -0.7}\) \\
    NGC6681\(^\ast\) & \(0.056\substack{+0.006 \\ -0.007}\)  & \(55\substack{+6 \\ -6}\)                   & \(6.5\substack{+0.7 \\ -0.7}\)  & \(8.5\)                                       & \(72.8\substack{+0.5 \\ -0.6}\) \\
    NGC6723          & \(0.05\substack{+0.07 \\ -0.04}\)     & \(90\substack{+120 \\ -70}\)                & \(8\substack{+9 \\ -6}\)        & \(12.2\)                                      & \(66\substack{+1 \\ -1}\) \\
    NGC6752\(^\ast\) & \(0.45\substack{+0.01 \\ -0.02}\)     & \(920\substack{+30 \\ -40}\)                & \(83\substack{+3 \\ -5}\)       & \(11.1\)                                      & \(67.5\substack{+0.4 \\ -0.4}\) \\
    NGC6779          & \(0.48\substack{+0.06 \\ -0.04}\)     & \(770\substack{+90 \\ -70}\)                & \(54\substack{+6 \\ -5}\)       & \(14.3\)                                      & \(53.9\substack{+0.9 \\ -1.0}\) \\
    NGC6809          & \(0.05\substack{+0.02 \\ -0.03}\)     & \(90\substack{+40 \\ -50}\)                 & \(6\substack{+3 \\ -4}\)        & \(14.3\)                                      & \(57\substack{+1 \\ -1}\) \\
    NGC7078\(^\ast\) & \(0.36\substack{+0.05 \\ -0.04}\)     & \(2200\substack{+400 \\ -300}\)             & \(290\substack{+40 \\ -30}\)    & \(7.9\)                                       & \(50.4\substack{+0.5 \\ -0.5}\) \\
    NGC7089          & \(0.4\substack{+0.1 \\ -0.1}\)        & \(2300\substack{+700 \\ -700}\)             & \(130\substack{+40 \\ -30}\)    & \(17.7\)                                      & \(40\substack{+2 \\ -2}\) \\
    NGC7099\(^\ast\) & \(0.042\substack{+0.003 \\ -0.002}\)  & \(56\substack{+5 \\ -3}\)                   & \(6.9\substack{+0.6 \\ -0.4}\)  & \(8.2\)                                       & \(53.2\substack{+0.7 \\ -0.6}\) \\
    \hline
    \end{tabular}
    \caption{Median and \(1\sigma\) credibility intervals of the BH
             mass fraction (\fbh), total mass in BHs (\({M}_{\mathrm{BH}}\)),
             total number of BHs (\({N}_{\mathrm{BH}}\)) and remnant mass
             fraction (\(f_{\mathrm{remn}}\)) in all clusters in our sample,
             as well as the median mean BH mass (\(\bar{m}_{\mathrm{BH}}\)).
             All clusters classified as core-collapsed in
             \citet{Trager1995} are denoted by an asterisk.
             Note that all uncertainties presented here represent only the
             statistical uncertainties on the fits, and likely underestimate
             the true uncertainties (see \Cref{sec:validation}).
             }
    \label{table:BH_results}
    \end{table*}

    \begin{landscape}
    \begin{figure}%[h]
        \centering
        \includegraphics[width=\linewidth]{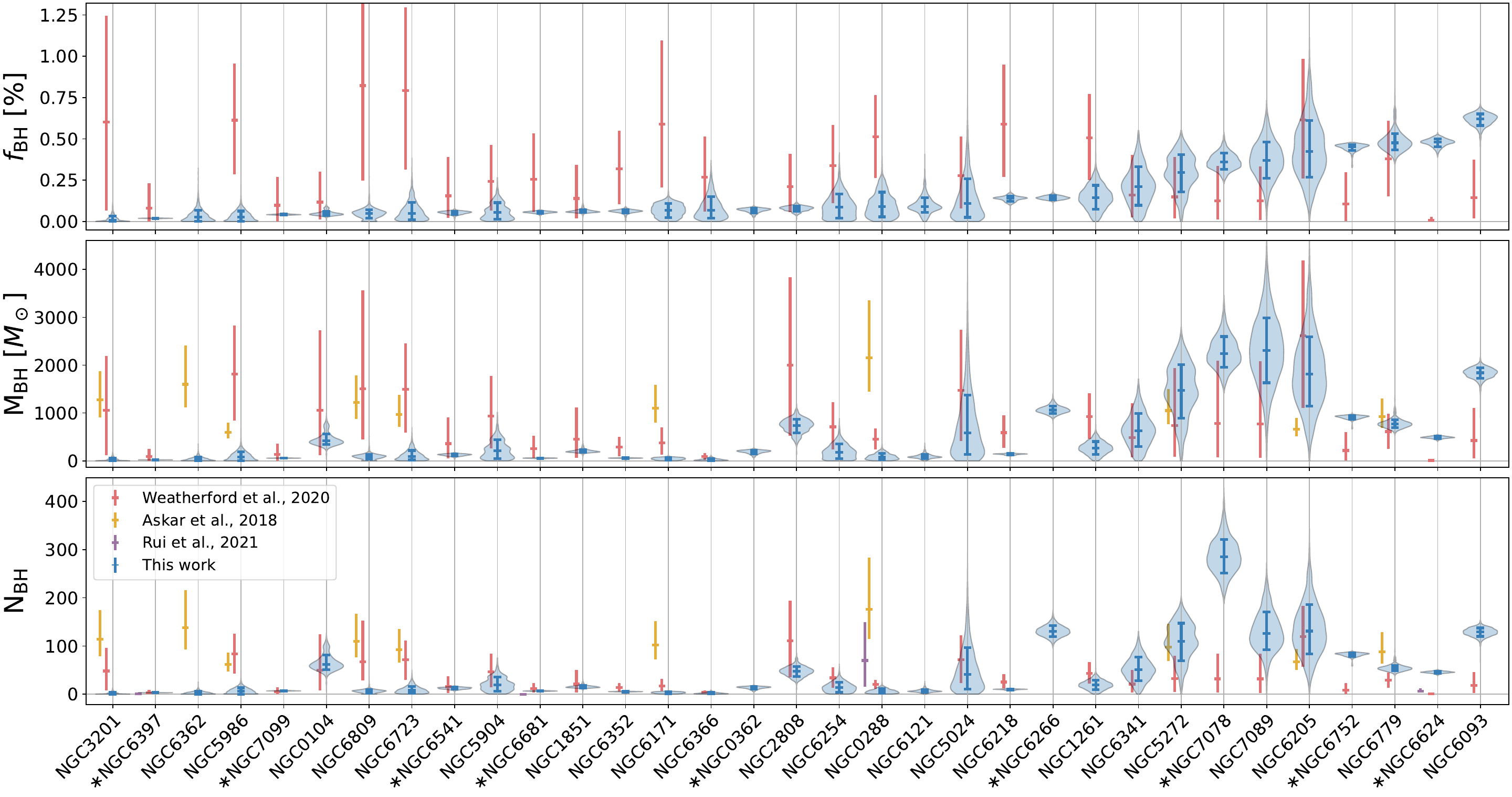}
        \caption{Violin plots (in blue) of the posterior probability
                 distribution of the mass fraction in BHs (upper panel), the
                 total mass in BHs (middle panel) and total number of BHs
                 (lower panel) in all clusters in our sample, except for
                 \NGC5139, which has a median total mass in BHs of
                 approximately
                 \SI{1.82e5}{\Msun} (\(\fbh \sim 5.9\) per cent),
                 and is excluded in order to better highlight the distributions
                 of the other clusters.
                 The median and \(1\sigma\) values are denoted by
                 the horizontal blue ticks within each distribution. These errors
                 include only the statistical uncertainties on our fits, and
                 thus are likely underestimated (see \Cref{sec:validation}).
                 Clusters are sorted based on the mass fraction in black holes.
                 All clusters classified as core-collapsed in \citet{Trager1995}
                 are denoted by an asterisk.
                 The median and \(1\sigma\) results are also shown for the
                 corresponding quantities in \citet{Weatherford2020} (red),
                 \citet{Askar2018} (orange) and \citet{Rui2021b} (purple),
                 for all clusters in common with our sample. Values from
                 \citet{Weatherford2020} are computed using
                 the median clustercentric mass segregation parameter
                 \(\Delta_{r50}\) (Table 1 of \citealt{Weatherford2020}),
                 and any necessary conversions between total mass and mass
                 fraction are computed using our total cluster mass estimates.
                 }
        \label{fig:BH_masses_nums}
    \end{figure}
    \end{landscape}

    \Cref{fig:BH_masses_nums} shows the posterior probability distributions of
    the mass fraction in BHs (\fbh), the total mass in BHs
    (\(M_{\mathrm{BH}}\)), and total number of BHs (\(N_{\mathrm{BH}}\))
    inferred from our best-fitting models of most clusters in our sample.
    The median and \(1\sigma\) credibility intervals of the distributions of
    these quantities, as well as the mean individual BH mass
    (\(\bar{m}_{\mathrm{BH}}\)) and the mass fraction in all dark remnants
    (\(f_{\mathrm{remn}}\)), are also presented in \Cref{table:BH_results}.
    \NGC5139 (\omegacen) is not included in \Cref{fig:BH_masses_nums}, due to
    the very large inferred amount of black holes
    (\(5.9\substack{+0.2 \\ -0.2}\) per cent),
    but it is discussed in more detail in \Cref{sec:clusters_of_interest}.

    A large number of the clusters are consistent, within \(2\sigma\), with
    harbouring little to no black holes, while the remainder possess, on
    average, at most a few thousand \Msun of stellar-mass black holes, with
    constituent individual BH masses ranging between \(\sim 5\) to
    \(15\ \Msun\).
    Other than \omegacen, it is clear that none of our models favour a very
    large population of black holes, with all clusters having a mass fraction
    in black holes between 0 and 1 per cent at the present day.

    It must be noted again that, as examined in \Cref{sec:validation}, the
    errors on all of these quantities represent only the statistical
    uncertainties, and in reality the uncertainties on a quantity like \fbh
    may be underestimated by a factor of \(\sim2.5\).

\subsection{Core-collapsed clusters}\label{sec:core_collapsed_clusters}

    %-----------------------------------------------------------------------
    % Core-collapsed clusters
    %-----------------------------------------------------------------------

    After all BHs are dynamically ejected, the visible core `collapses'
    \citep{Breen2013a}.
    Globular clusters having undergone core collapse are typically defined
    based on the shape of their central density profiles, with
    core-collapsed clusters showing a power-law density profile increasing all
    the way to their centres, while non core-collapsed clusters possess larger,
    isothermal cores with a flat central density profile
    \citep[e.g.][]{Djorgovski1986,Trager1995}.
    Core-collapsed clusters are expected to contain very few, if any, black
    holes at the present day \citep{Breen2013a,Breen2013b,Kremer2020a}.
    In GCs with a population of stellar-mass BHs,
    core collapse occurs within the BH subsystem but, due to the
    efficient heat transfer from BHs to stars, the visible core will actually
    remain large (relative to \(r_{\mathrm{h}}\)).
    The presence of BHs in a cluster may thus play a large role in explaining
    the observed population of core-collapsed Milky Way GCs, which, given the
    ages of most clusters, is smaller than would be expected when considering
    only stellar binaries as the sole mechanism delaying core collapse.
    It is not until almost all BHs \citep[and the last BH binary;][]{Hurley2007}
    are ejected that a cluster core will collapse and exhibit the defining
    power-law central density profile \citep{Chatterjee2013,Kremer2020a}.
    Almost all GCs have likely reached a state of balanced evolution
    \citep{Henon1961, Gieles2011} but, due to BHs, only a minority  of GCs
    (\(\sim20\%\)) \textit{appear} to be post-collapse.

    The nine clusters in our sample defined as core-collapsed in
    \citet{Trager1995} (\NGC362, \NGC6266, \NGC6397, \NGC6541, \NGC6624,
    \NGC6681, \NGC6752, \NGC7078, \NGC7099) are denoted in
    \Cref{fig:BH_masses_nums,table:BH_results} by an asterisk.
    Our best-fitting models of some of these clusters clearly favour
    a non-negligible amount of mass in black holes, however, as discussed below,
    we suspect the results are unreliable for these few clusters.

    Part of this discrepancy between the theoretical expectations and the
    inferred BH populations for some core-collapsed clusters may arise simply
    due to the limitations of the \limepy models themselves. 
    \limepy models, by definition, possess an isothermal core,
    characterized by a flat inner density profile, which is incompatible with
    the central cusp of core-collapsed clusters (see also Section 3.1.4 of
    \citealt{Gieles2015}).
    As such, our models may struggle to accurately
    capture the inner density profiles of these clusters right up to the centre.
    Indeed this divergence can be seen in the profiles of the core-collapsed
    clusters with for which we infer substantial BH populations from our
    best-fitting models, which tend to underestimate the amount of stars within
    a very small distance from the centre
    (typically \(\sim \SI{0.1}{\pc}\) in these clusters).
    While, in most core-collapsed clusters, the models may be able to provide a
    satisfactory fit to the available density data simply by
    having a sufficiently small core (below the radial reach of
    the data), in some clusters the shape of other datasets (such as the
    mass functions) may require a larger core, and cause the fitting procedure
    to sacrifice the quality of the fit to the central density profiles.

\begin{figure}%[h]
        \centering
        \includegraphics[width=\linewidth]{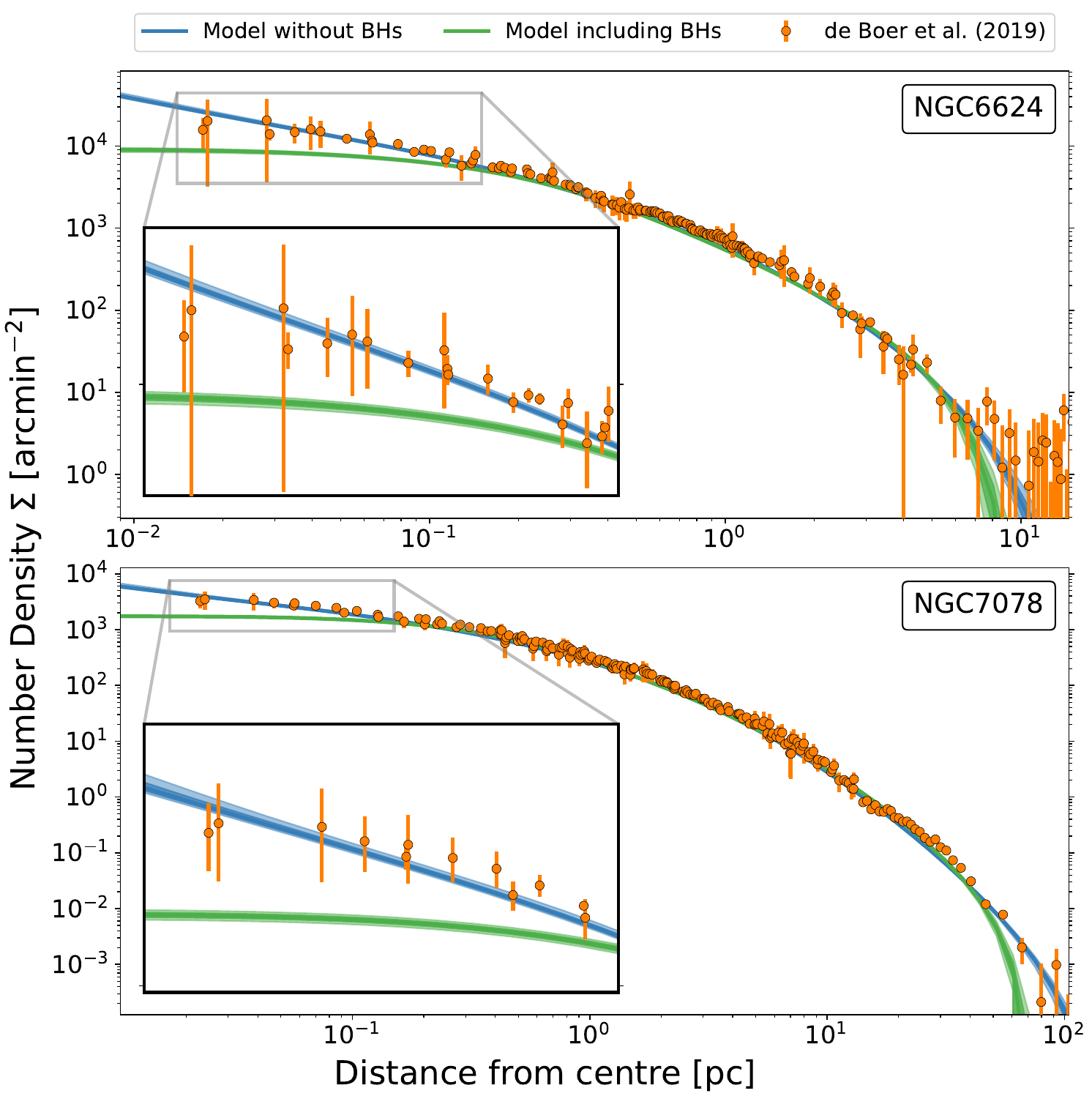}
        \caption{The number density profiles of the best-fitting models of
                 \NGC6624 and \NGC7078, with and without
                 allowing for a population of BHs. The number density data used
                 to constrain the models is shown by the orange circles.
                 Background levels have been subtracted from all data.
                 Inset frames show a zoomed view on the model profiles near the
                 cluster cores, to showcase the differences between the sets of
                 models.}
        \label{fig:CC_numdens_profiles}
    \end{figure}

    In order to investigate these systems further, models were also fit to
    these clusters, in the same fashion as before, but with the amount of
    retained black holes at the present day now fixed to 0
    (by fixing the \(\mathrm{BH}_{\mathrm{ret}}\) parameter to 0 per cent).
    As might be expected, the most immediately noticeable change to the model
    fits is in the number density profiles.
    Shown in \Cref{fig:CC_numdens_profiles} are examples of the changes
    between the sets of models for \NGC6624 and \NGC7078, the two core collapsed
    clusters in our sample with the highest inferred mass fraction in BHs
    (\(\fbh \sim 0.5\) per cent) and the only two where significant differences
    can be seen between the models with and without BHs.
    In the other core-collapsed clusters, which favour smaller or
    negligible BH populations, the fits do not change noticeably. Likely, as
    suggested above, the cores of these models are small enough that even
    though the models are not truly core collapsed, they are still able to
    represent the observed density profiles.
    In the case of \NGC6624 and \NGC7078, \Cref{fig:CC_numdens_profiles}
    clearly shows that models without BHs \textit{are} able to also reproduce
    the central density profiles, and would provide a much better fit in that
    regime, however, the models containing BHs have a higher overall likelihood,
    due to the fits to other datasets, which cannot be reproduced as well by
    the models without BHs.
    Overall, it is clear that caution must be applied when
    attempting to fit core-collapsed clusters using \limepy models.

    The original models (with BHs) are used throughout this paper and in
    all discussion of BHs, however the results for these core-collapsed clusters
    should be regarded with great caution, especially when a large
    population of BHs is inferred.
    All core-collapsed clusters are noted in the figures and tables by an
    asterisk or a red outline.
    As the larger inferred BH populations of some of these clusters
    are most likely artificial, the models introduced here, fixed to 0 BHs,
    were used in \paperI to avoid any impacts on the inferred mass function
    slopes (though these were found to be minimal).

\subsection{Relationships between BH population and other parameters}
\label{sec:relationships_between_bh_population_and_other_parameters}

    We next examine how the population of black holes and other remnants in
    our cluster models correlate with various related parameters.

    %-----------------------------------------------------------------------
    % Relationship with BH forming parameters and the final BHs
    %-----------------------------------------------------------------------

    \begin{figure}%[h]
        \centering
        \includegraphics[width=\linewidth]{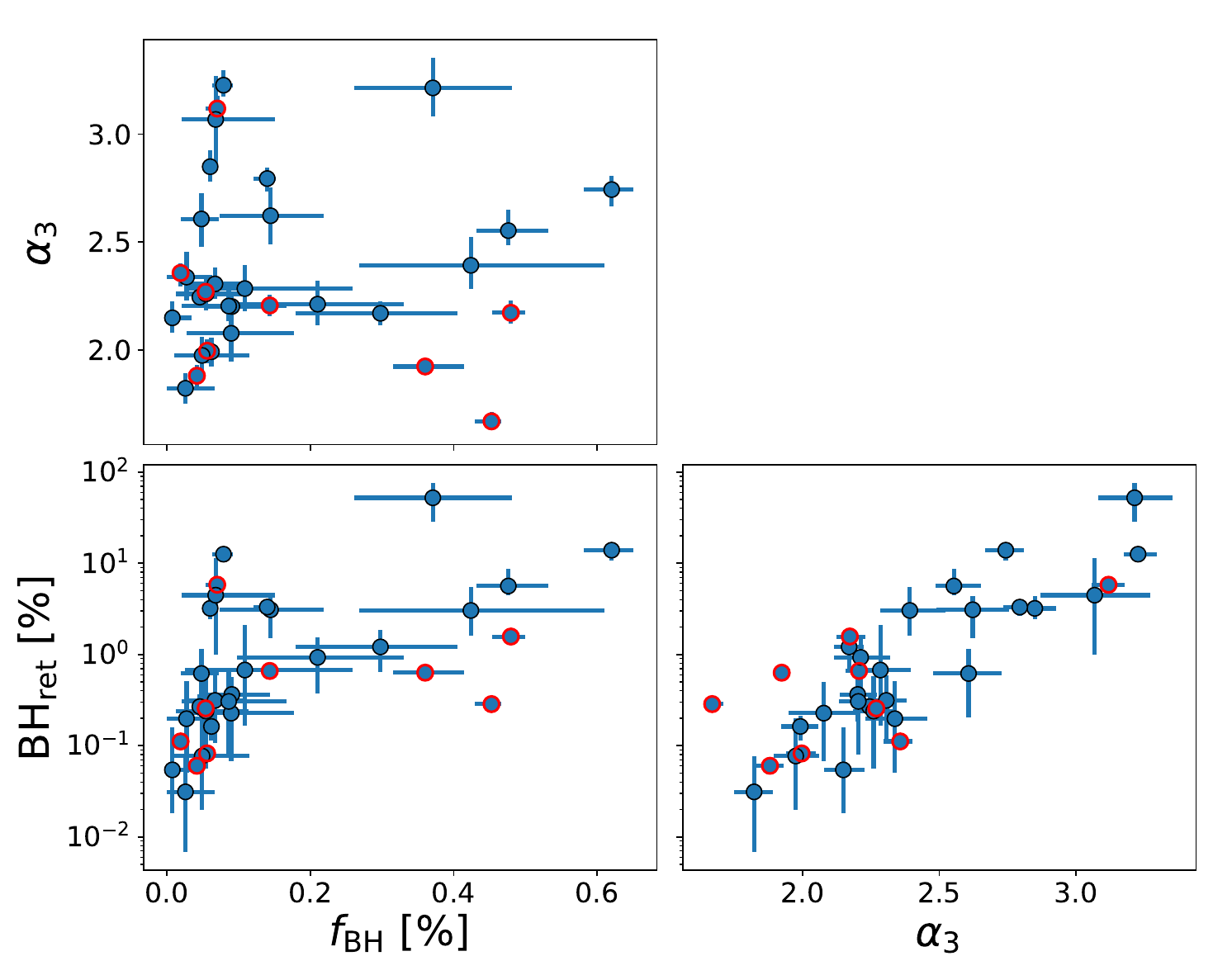}
        \caption{Relations between the high-mass initial mass function
                 exponent (\(\alpha_3\)), the black hole retention
                 fraction parameter (\(\mathrm{BH}_{\mathrm{ret}}\)) and the
                 mass fraction in black holes (\fbh) for all clusters, except
                 for \NGC{5139}, which has a high \(\mathrm{BH}_{\mathrm{ret}}\)
                 of \(\sim 35\substack{+9 \\ -3}\) per cent and a more typical
                 \(\alpha_3\) value of \(\sim 2.28\substack{+0.08 \\ -0.02}\).
                 All core-collapsed clusters, whose inferred black hole
                 populations may not be physical, are highlighted by a
                 red outline.
                 }
        \label{fig:a3_BHret_f_BH}
    \end{figure}

    \Cref{fig:a3_BHret_f_BH} shows the relationships between
    the high-mass initial mass function exponent \(\alpha_3\)
    (\(\geq \SI{1}{\Msun}\)) from \paperI, the black hole
    retention fraction \(\mathrm{BH}_{\mathrm{ret}}\) and the BH mass fraction
    \fbh.
    This serves to demonstrate the role of the \(\mathrm{BH}_{\mathrm{ret}}\)
    parameter, which is not \textit{directly} proportional to the number of BHs.
    At high values of \(\alpha_3\) (i.e. steeper slopes), only a small number
    of black holes can be formed initially from the IMF, and a higher retention
    fraction is required to maintain any amount of black hole mass at the
    present day.
    The top panel of \Cref{fig:a3_BHret_f_BH} shows that
    no clear correlation is present between \(\alpha_3\) and the mass
    fraction in BHs at the present day. The visible pattern instead relates
    to the relationship with \(\mathrm{BH}_{\mathrm{ret}}\); to end up with
    similar mass fractions in BHs, clusters with higher \(\alpha_3\) values
    produce fewer BHs initially while clusters with low \(\alpha_3\) values
    produce more BHs initially but retain few.
    This degeneracy can be clearly seen in the rightmost panel.
    This of course does not imply a causal relationship between BH mass
    fraction and either \(\alpha_3\) or \(\mathrm{BH}_{\mathrm{ret}}\), but
    merely helps to explain the distribution of \(\mathrm{BH}_{\mathrm{ret}}\).
    For a given cluster there is generally not a direct degeneracy between
    \(\alpha_3\) and \(\mathrm{BH}_{\mathrm{ret}}\) apparent in the
    posterior probability distributions of the parameters, as the shape of the
    IMF and retention factor also strongly impact the amount of other massive
    stars and remnants formed, which may be strongly constrained by the
    observations.

    %-----------------------------------------------------------------------
    % Relationship with mass segregation
    %-----------------------------------------------------------------------

    \begin{figure}%[h]
        \centering
        \includegraphics[width=\linewidth]{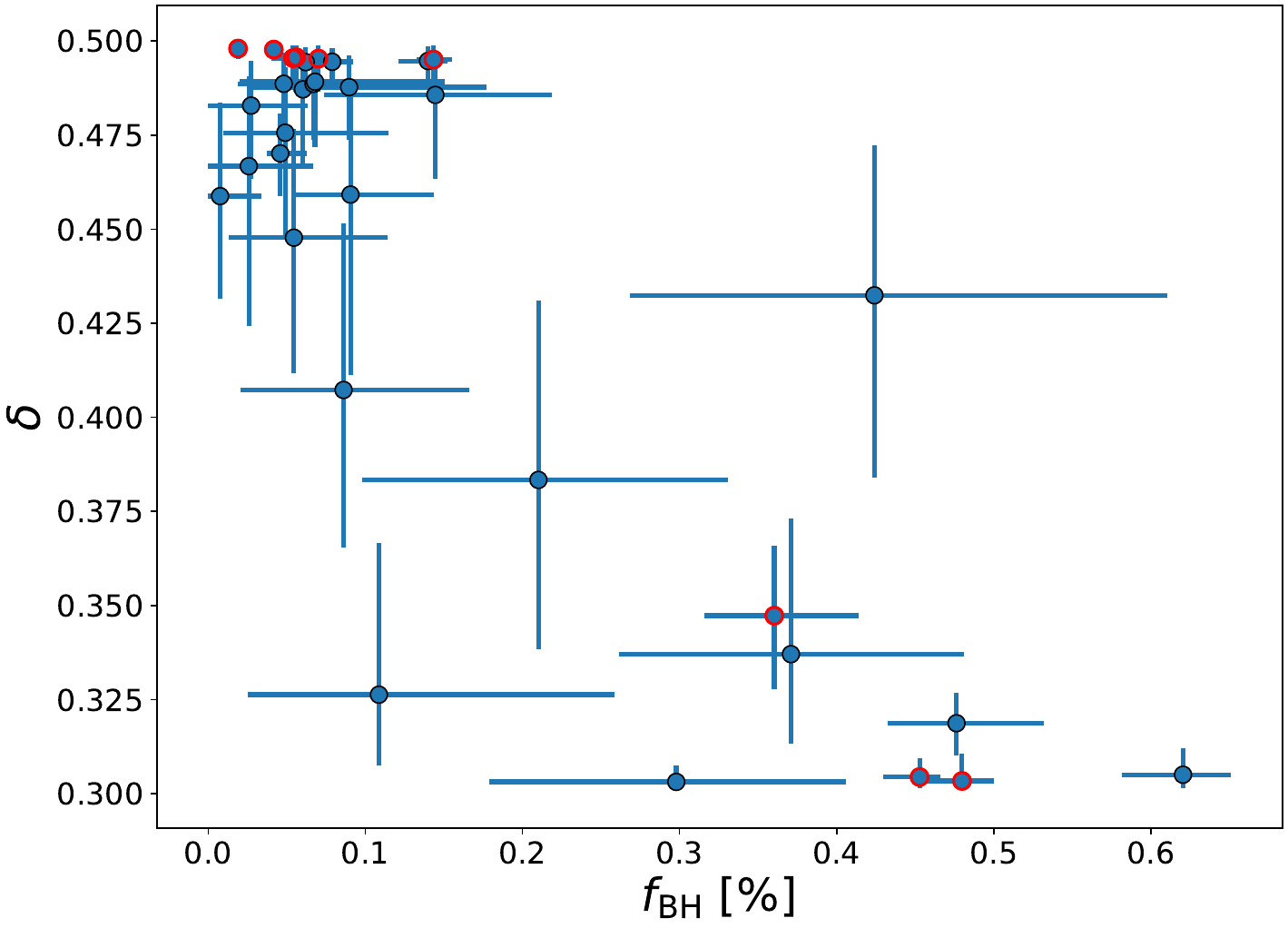}
        \caption{Relation between the mass-dependant velocity scale (\(\delta\))
                 and the mass fraction in black holes
                 for all clusters, except \NGC5139.
                 All core-collapsed clusters, whose inferred black hole
                 populations may not be physical, are highlighted by a
                 red outline.
                 }
        \label{fig:delta_f_BH}
    \end{figure}

    We do find an interesting relationship between the mass fraction of BHs
    and the parameter \(\delta\), which sets the mass-dependence of the velocity
    scale and acts as a proxy of mass segregation, as shown in
    \Cref{fig:delta_f_BH}. Clusters with little to no mass in BHs
    tend to converge near values of \(\delta \sim 0.4 \rm{-} 0.5\), which is
    typical of evolved and mass-segregated clusters, whereas the
    clusters with more substantial populations of black holes congregate
    closer to the lower bound of 0.3. This is in agreement with the
    study of \citet{Peuten2017}, who find, by comparing \limepy models against
    \Nbody models with and without black holes, that the majority of
    mass-segregated clusters should converge to a value of \(\sim0.5\), but
    also show that, in models with a significant population of black holes,
    the degree of mass segregation as traced by the parameter \(\delta\) may
    be suppressed.

% !TEX root = ./paper.tex

%---------------------------------------------------------------------------
\section{Discussion}\label{sec:discussion}
%---------------------------------------------------------------------------

%---------------------------------------------------------------------------
\subsection{The evolution of clusters and their BH populations}
%---------------------------------------------------------------------------

    The co-evolution of a cluster and its BH population depends directly
    on the BH mass fraction \fbh and the initial density of the cluster.
    Very early on in the lifetime of a cluster, \fbh will rapidly increase, as
    stellar mass is lost due to stellar evolution and BHs are formed.
    This initial population of BHs, and \fbh, depends primarily on the cluster
    metallicity, the shape of the IMF\, and the velocity distribution of natal
    kicks imparted on BHs when they form.
    The BHs initially retained (after kicks) will rapidly segregate to the
    cluster core. During the proceeding expansionary phase, BHs will be ejected
    from the core due to dynamical BH-BH interactions, while the lowest mass
    stars, on the cluster periphery, will be
    the most affected by tidal losses, leading to a decrease in \fbh
    \citep{Breen2013b,Gieles2021}.
    The amount that \fbh decreases during this stage is dependent on the
    initial cluster density, relative to the tidal density. Higher density
    clusters will eject a higher proportion of the initially formed BHs
    \citep{Banerjee2011}.

    Once the cluster becomes tidally limited, the remaining \fbh determines the
    subsequent evolution of the cluster mass.
    Theory \citep{Breen2013b} and \Nbody modelling \citep{Gieles2023} suggests
    that there exists a critical BH mass fraction \(f_{\mathrm{crit}}\), at
    which the dynamical losses of BHs (relative to the total cluster mass loss)
    will exactly equal \fbh, causing \fbh to remain constant until the
    dissolution of the cluster. Given an \fbh greater(less) than this critical
    fraction, tidal stellar mass losses will be greater(less) than the
    dynamical ejections of BHs from the core. Clusters above the
    critical fraction will progress towards 100 per cent BHs, while clusters
    below will eject most of the BHs and begin to evolve similarly to a 0 BH
    system.
    This critical fraction was determined to be \(\sim\) 10 per cent
    \citep{Breen2013b} for idealized single-mass clusters filling their tidal
    radius, however recent \Nbody modelling placed it closer to 2.5 per cent
    for clusters with a full mass spectrum \citep{Gieles2021}.

    In our sample, while there is a range of values between individual
    clusters, we find that nearly all clusters have a \fbh below 1 per cent
    (except for \omegacen).
    \citet{Gieles2023} modelled the Milky Way GC population using a
    parametrization for the mass-loss rate from GCs with BHs, assuming
    that the initial GC half-mass density is 30 times higher than Roche
    filling. Under this assumption, they predicted that the majority of GCs
    should now have \(\fbh \sim 2\) per cent.
    Our empirical results are lower than this,
    which could indicate that their initial densities are not high enough.
    However, the authors do note that the constant density (relative to
    the tidal density) they have used for all GCs is only meant to approximate
    the \textit{average} density of GCs.
    In reality, a distribution of ratios of initial densities over the tidal
    densities must exist, which would lead to both a population of GCs that can
    eject all their BHs (the highest density GCs) and a population of
    extended GCs (like the Palomar GCs) which fill the Roche volume (the lowest
    density GCs) \citep{Baumgardt2010}.
    Lower density GCs would also have a higher mass-loss rate, and therefore
    there may exist a survival bias in the clusters visible at the present-day,
    biasing our sample.
    It must also be noted that the initial \fbh is highly dependent on the
    IMF of the clusters. The IMF of our clusters is likely
    bottom-light compared to a Kroupa IMF
    \citep[see \paperI and][]{Baumgardt2023}. This would lead to an increased
    initial \fbh, and thus require an even higher density in order to reach
    the critical fraction.

    As the \fbh we see in our clusters is well below the critical mass
    fraction noted by \citet{Gieles2021}, it is clear that the \fbh in these
    clusters must have been below this critical amount if/when they became
    tidally limited. The clusters can thus be expected to continue to evolve
    towards BH free clusters.

    % ---

    Early theoretical expectations of the behaviour of GCs and their compact
    BH subsystem suggested that the subsystem would become dynamically decoupled
    from the rest of the stars in the cluster, succumbing to the
    \textit{Spitzer Instability} \citep{Spitzer1969,Spitzer1987}, and rapidly
    ejecting nearly all BHs.
    However, more recent modelling work has shown that clusters do not
    necessarily become Spitzer unstable \citep{Morscher2013,Morscher2015}.
    In our models, a large fraction (\(24 / 34\)) of the clusters remain Spitzer
    stable (according to the classification given in \citealt{Spitzer1987}),
    consisting mostly of clusters with \(\fbh \lesssim 0.1\) per cent.
    \citet{Breen2013a}, examining an idealized, two-component model of clusters,
    suggested a relationship between the mass fraction in BHs and the relative
    size of the BH subsystem, proportional to \(\fbh^{3/2}\) for Spitzer
    unstable systems and \(\fbh\) for stable systems.
    While it is difficult to compare directly with the fits on this relation
    by their simple models, they do predict a BH subsystem size of
    \(r_{h,\mathrm{BH}}/ r_h \sim 0.04\) near \(\fbh \sim 0.1\) per cent, which
    is similar to the bulk of our clusters.

    % ---

    \citet{Baumgardt2023}, examining a suite of \Nbody models compared to
    observations, demonstrated that the trend seen between the global low-mass
    stellar mass function slopes, as derived from
    observations, and the dynamical age of clusters could be reproduced only
    by models with a maximum BH retention fraction of 30 per cent (immediately
    after natal kicks), ruling out a high initial retention rate.
    Their models with higher initial BH retention fractions (which suppresses
    mass segregation and the preferential loss of low-mass stars) cannot
    reproduce the trend since they are not able to produce clusters strongly
    depleted in low-mass stars.
    After the subsequent BH hardening and dynamical ejections, they estimate
    an average surviving BH mass fraction of \(\fbh \sim 0.03 - 0.1\) per cent
    at the present day, in generally good agreement with many of our clusters.

    It should be noted that any analysis of our results with respect
    to the complete population of Milky Way GCs could be biased by the choice
    of clusters examined.
    The sample of clusters chosen was limited primarily by the availability of
    good quality data, especially mass function depth and radial coverage,
    requiring adequate deep \HST photometry. These criteria bias our sample
    somewhat towards more massive and nearby clusters.
    In comparison to the overall population of GCs
    \citep[as given in][2010 edition]{Harris}, our chosen clusters tend to
    be slightly more massive, have a smaller core radius and a lower
    galactocentric radius.This could indeed bias our sample slightly towards
    clusters with lower \fbh,
    as they do not contain many low-density or `fluffy' outer halo GCs,
    which may have much higher \fbh \citep{Gieles2021}.
    Correspondingly, as was noted in \paperI, our sample is deficient
    in the most metal-rich clusters, which are expected to produce notably
    different remnant populations.

    % ---

    \begin{figure}%[h]`'
        \centering
        \includegraphics[width=\linewidth]{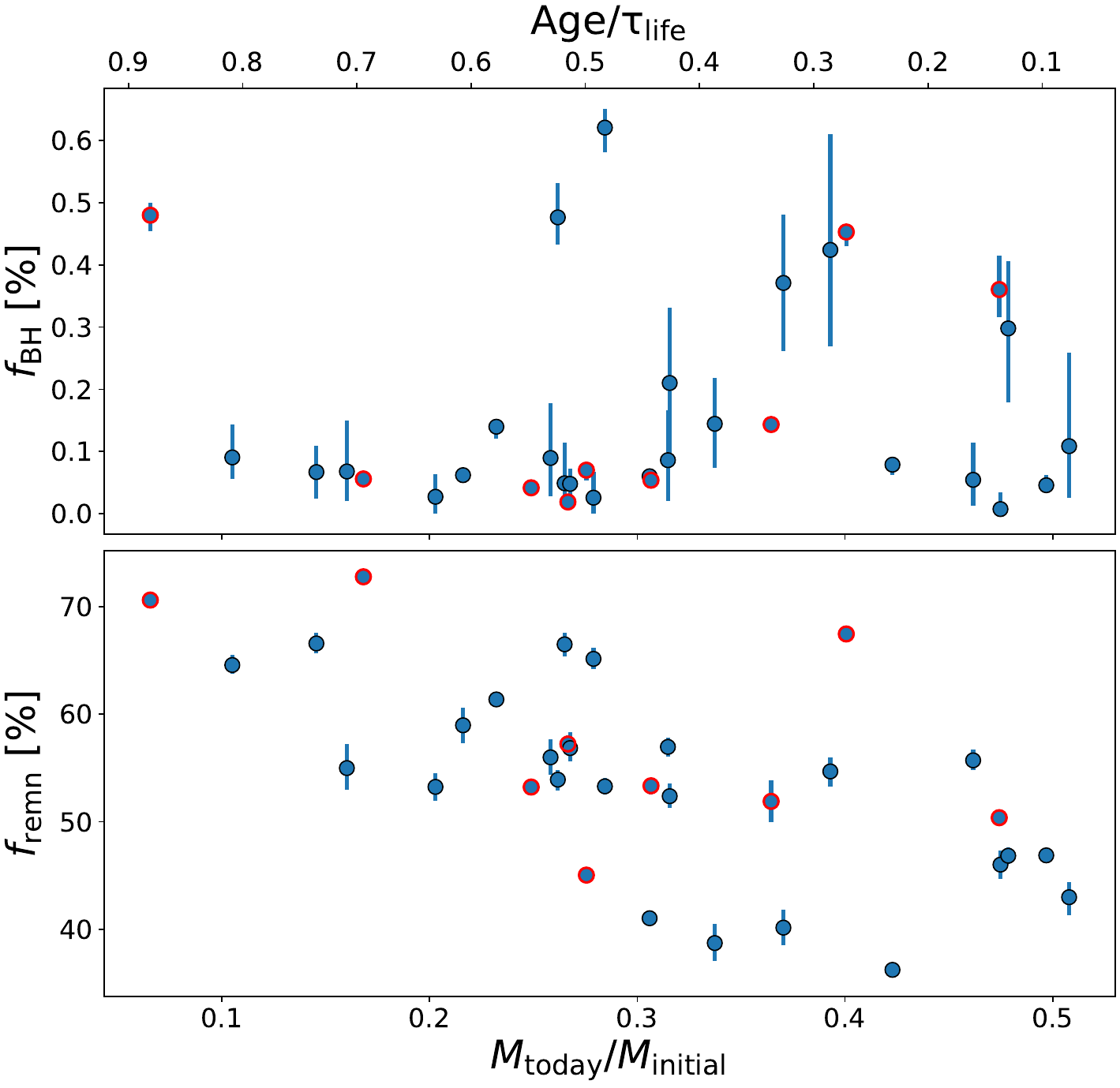}
        \caption{Relations between mass fraction in BHs (top panel)  and
                 mass fraction in all dark remnants (bottom panel) with the
                 remaining mass fraction for all clusters, except for \NGC{5139}
                 (\(M_{\rm{today}}/M_{\rm{initial}} = 0.50\), \(\fbh = 5.9\)
                 per cent, \(f_{\mathrm{remn}} = 50.3\) per cent).
                 All core-collapsed clusters, whose inferred BH populations
                 may be unreliable, are highlighted in red.
                 }
        \label{fig:BH_mass_frem_mu_dyn}
    \end{figure}

    We can also examine how the populations of remnants in our clusters
    relate to their dynamical age.
    \Cref{fig:BH_mass_frem_mu_dyn} shows the relationship between the
    fraction of mass in BHs and the fraction of the cluster
    mass in all remnants (WD, NS and BHs), against the dynamical age of
    the clusters, estimated based on the \textit{remaining mass fraction}, as
    was described in \paperI:
    \begin{equation}\label{eq:remaining_mass_frac}
        \frac{M_{\mathrm{today}}}{M_{\mathrm{initial}}} = 
            0.55 \times \left(1- \frac{\mathrm{Age}}{\tau_{\mathrm{life}}} \right),
    \end{equation}
    where the factor 0.55 reflects the typically assumed mass loss from
    stellar evolution of \(\sim 45\) per cent of the initial cluster mass
    in the first Gyr of a cluster's evolution and the dissolution time
    \(\tau_{\mathrm{life}}\) represents the estimated total lifetime of the
    cluster and is taken from \citet{Baumgardt2023}.

    While we can see here that clusters with substantial populations of
    BHs tend to be less evolved, there is no strong correlation between the
    BH mass fraction and the dynamical age of the clusters, likely indicative
    of their different initial conditions and the effects they would have on
    the evolution of the BH population over time, as discussed above.

%---------------------------------------------------------------------------
% \subsection{Remnant mass fractions}
%---------------------------------------------------------------------------

    The evolution of the \textit{remnant} mass fraction, which includes all
    types of stellar remnants, shows a stronger relationship with the dynamical
    age of the clusters, as might be expected,
    and has been previously reported by \citet{Sollima2017}.
    As a cluster evolves and loses mass, the mass lost is
    preferentially in the form of lower-mass stars from the outer parts of the
    cluster, rather than the heavier remnants,
    and as such the fraction of mass in remnants
    should increase as the cluster's low-mass stars are depleted.
    Interestingly, some of the most dynamically evolved clusters have around
    70 per cent of their mass in dark remnants at the present day, something
    worth bearing in mind when interpreting the mass-to-light ratios and
    inferring masses of unresolved GCs in distant galaxies.

    These relatively high remnant mass fractions are in good agreement with the
    results of \citet{Sollima2017}, for the clusters in common, although it
    should be noted that these authors adopt a simpler prescription for the
    mass function of remnants and a fixed, canonical high-mass IMF, unlike
    the flexibility allowed in our models for both of these quantities.
    The \Nbody models of \citet{Baumgardt2003} also showed that evolved
    clusters can consist of up to nearly 70 per cent WDs, by mass, in line with
    our results.
    As detailed in \paperI and \citet{Baumgardt2023}, our dynamical masses are
    also in excellent agreement with other recent works, and therefore the
    cluster mass-to-light ratios implied by our results are in keeping with the
    recent literature \citep[e.g.][]{Baumgardt2020}. These mass-to-light ratios
    are confined to a narrow range near \(M/L_V\sim 2\) and have been shown to
    be consistent with the the mass-to-light ratios predicted by stellar
    population models once the depletion of low-mass stars is taken into
    account \citep{Baumgardt2020}. Interestingly, the higher remnant mass
    fractions compensate the lack of low-mass stars to maintain the
    mass-to-light ratios within this narrow range, a finding consistent with
    the results of \citet{Sollima2012}.

%---------------------------------------------------------------------------
\subsection{Comparison with literature results}
\label{sec:comparison_with_literature_results}
%---------------------------------------------------------------------------

    %-----------------------------------------------------------------------
    % Comparison with Monte Carlo models
    %-----------------------------------------------------------------------

    Also shown in \Cref{fig:BH_masses_nums} are comparisons with the
    distribution of BH mass fraction, total mass in BHs, and/or number of BHs
    in our models with those of \citet{Askar2018}, \citet{Weatherford2020}
    and \citet{Rui2021a}.
    In order to estimate the BH mass fraction in a number of Milky Way GCs,
    \citet{Weatherford2020} compared the amount of visible mass segregation in
    these clusters to the anti-correlation found in \citet{Weatherford2018}
    between the degree of mass segregation in a cluster and its BH
    population in the \textit{Cluster Monte Carlo} (CMC) catalogue of models.
    In similar fashion to their analysis, we also scale
    their computed estimates of \fbh
    (based on the median clustercentric mass segregation parameter
    \(\Delta_{r50}\)) by the total cluster mass determined by our models in
    order to compare the total mass in BHs.
    \citet{Askar2018} predicted the amount of BHs in a number of Milky Way
    GCs based on the correlations found in \citet{ArcaSedda2018}
    between the density of inner BH-subsystems and the central surface
    brightness of the clusters in the \textit{Monte Carlo Cluster Simulator}
    (MOCCA) survey database.
    A somewhat analogous analysis may be found in \citet{Rui2021a}, who
    matched the surface brightness and velocity dispersion profiles of 26 Milky
    Way GCs to a grid of CMC models and explored seven in more detail (see
    \Cref{sec:validation} for more information). The number of
    BHs reported for the three clusters in common with \citet{Rui2021a} is
    shown in \Cref{fig:BH_masses_nums}.
    Further analysis and comparison with other studies of interesting
    individual clusters is presented below in \Cref{sec:clusters_of_interest}

    \begin{figure}%[h]
        \centering
        \includegraphics[width=\linewidth]{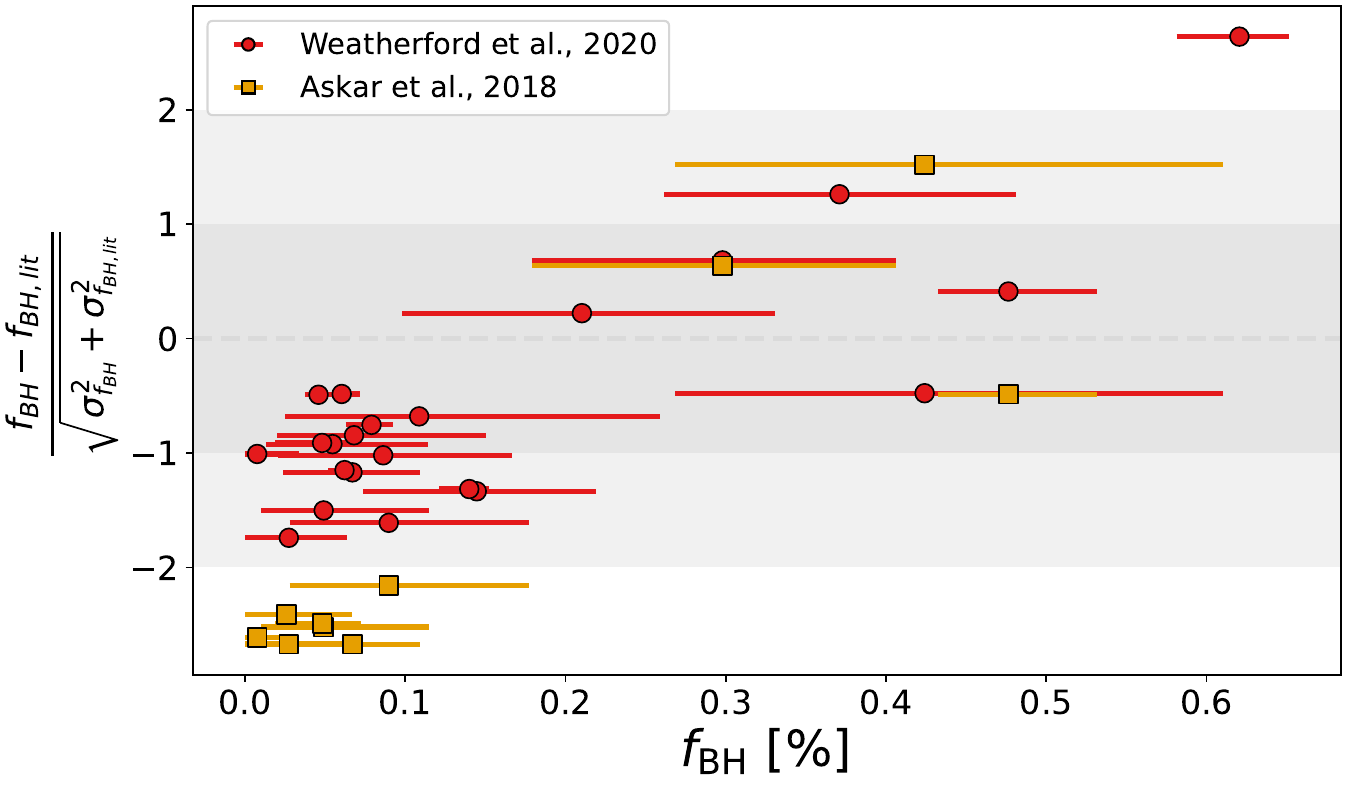}
        \caption{The (error weighted) residuals of the mass fraction in BHs
                 between literature sources (\citet{Weatherford2020} in
                 red and \citet{Askar2018} in orange) and our models,
                 with respect to the mass fraction in BHs of our models,
                 for all clusters overlapping with our sample, excepting any
                 core-collapsed clusters.
                 \citet{Weatherford2020} values are computed using
                 the median clustercentric mass segregation parameter
                 \(\Delta_{r50}\) (Table 1 of \citealt{Weatherford2020}).
                 Conversions between the total BH masses presented by
                 \citet{Askar2018} and BH mass fractions are computed using
                 our total cluster mass estimates.
                 }
        \label{fig:f_BH_rescomp}
    \end{figure}

    The majority of our clusters agree well, within \(2\sigma\),
    with the amount of mass in BHs estimated in these studies, however
    notable discrepancies can be seen between individual clusters.

    In \Cref{fig:f_BH_rescomp}, the residuals (normalized by the combined
    \(1 \sigma\) uncertainties) between our BH mass fraction results and
    those of the literature sources discussed above are shown.
    We can see a clear trend showing that for clusters where we infer small
    BH populations, we tend to predict fewer BHs than other studies, while for
    clusters where we infer larger BH populations, we tend to predict more BHs
    than previously reported in the literature.
    This is especially pronounced in the comparisons
    with \citet{Askar2018}, where the clusters with smaller BH populations
    do not even agree within \(2\sigma\). 
    In other words, the distributions of total BH masses between clusters
    predicted by \citet{Askar2018} and \citet{Weatherford2020} are somewhat
    flatter (i.e. less variation in BH mass fraction across the sample) than
    our results.
    Despite the differences between specific clusters in our samples, we
    do agree with the overall conclusion that, in general, the mass fraction
    of BHs retained in clusters at the present day is small, between 0 and 1
    per cent.

    Our analysis of the BH populations of individual clusters may be
    more robust than many of these literature results, which rely
    on general correlations between models with only a few varied initial
    parameters, and are fit on only a single observed property (mass
    segregation between three stellar populations for \citet{Weatherford2020}
    and the central surface luminosity for \citet{Askar2018}),
    whereas we self-consistently include the
    effect of BHs in our fits of numerous cluster observables with many
    free parameters.
    In particular, as noted in \citet{Weatherford2020}, the correlations of
    \citet{Askar2018} rely on a number of chained parametric fits, which may
    bias the final values, and the models used to construct the correlations
    in \citet{ArcaSedda2018} exclude those with \(N_{\mathrm{BH}} < 15\), which
    may lead to overpredictions in their inferred numbers of BHs.
    \citet{Rui2021a} matched available CMC model snapshots to profiles of
    surface brightness and velocity dispersion observations, but this study is
    based on a limited grid of models with only a few varied parameters.
    We are thus able to, in most clusters, place tighter constraints on
    the mass in BHs through our fits.
    It should, however, be noted again that our uncertainties
    account solely for the statistical uncertainties on the parameter fits and
    could thus be underestimated by a factor of up to \(\sim2.5\)
    (see \Cref{sec:validation}).
    In addition, there are a number of astrophysical processes that could
    possibly have a small effect on the inferred amount of which we do not
    model, such as binaries and cluster rotation.

    Another possible major source of differences between these results is
    the (initial) mass function formulation, which was identified by
    \citet{Weatherford2020} and \citet{Rui2021a} as a potential source of
    unexplored uncertainty in their analysis. The freedom in the
    shape of our (initial) mass function, on a per-cluster basis, allows us to
    best explore the population of BHs and other heavy remnants, as well as
    their relative abundance compared to lower-mass stars.
    The generally more bottom-light, low-mass mass
    function slopes found in our fits (see \paperI) and in similar analyses
    \citep{Baumgardt2023}, compared to the canonical Kroupa IMF assumed by
    \citet{Weatherford2020} and others, may for example affect the mass
    segregation correlation found in \citet{Weatherford2018} and thus the
    amount of mass in BHs found by \citet{Weatherford2020}.
    However, the exact effects of such a bottom-light IMF
    on cluster evolution, and the evolution of BHs within the
    clusters, remain to be further explored.

    We can also place our BH results in the context of a number of studies
    that have searched for the presence of an intermediate-mass
    black hole (IMBH) in various Milky Way GCs, based on a number of indirect
    dynamical inferences \citep[e.g.][]{Noyola2008,McNamara2012,Kiziltan2017,
    Perera2017a}. All of these reported IMBH candidates have been controversial,
    and in many cases have been rebutted, either by the introduction of
    improved data, or by other, more plausible, physical interpretations of the
    data \citep[e.g.][]{vanderMarel2010,Zocchi2017,Zocchi2019,Gieles2018,
    Baumgardt2019b}.
    Our models are able to reproduce well the observables of all clusters in
    our sample without the need for any central IMBH.
    It should, however, be noted that it is currently not possible to
    self-consistently include an IMBH in our models to compare directly and
    to explore any partial degeneracy possible between a central IMBH and a
    central concentration of stellar-mass BHs \citep[e.g.][]{Lutzgendorf2013}.
    It has also been shown that an IMBH with a mass fraction below
    \(\sim 2\) per cent of the cluster mass may produce similar dynamical
    effects as a population of stellar-mass BHs \citep{Aros2023}.
    Therefore we cannot preclude their existence with certainty with
    our models, given these degeneracies.
    Some of the clusters previously claimed to host an IMBH are discussed in
    more detail in \Cref{sec:clusters_of_interest} below.

%---------------------------------------------------------------------------
\subsection{Comments on individual clusters}\label{sec:clusters_of_interest}
%---------------------------------------------------------------------------

    We now compare our results with other dynamical studies of BHs in GCs,
    based on a number of different methods, in certain particularly interesting
    clusters.

\subsubsection{\NGC5139}

    \NGC5139, or \omegacen, is the most massive Milky
    Way GC, and stands apart from the population of Milky Way GCs due to its
    mass and stellar populations (\citealp[e.g.][2010 edition]{Harris};
    \citealp{Baumgardt2019a}).
    It has been suggested that \omegacen may not be a classical globular
    cluster, but rather the possible
    remnant nuclear star cluster of an accreted and disrupted dwarf galaxy
    \citep[e.g.][]{Meza2005}. It has also been hypothesized to harbour
    an elusive IMBH \citep{Noyola2008,vanderMarel2010}.

    While our models are able to fit the large amount of
    available data very well, it does not appear in many of the figures
    above given its significantly larger inferred mass fraction in BHs.
    Our fits favour a mass fraction in BHs of
    \(5.9\substack{+0.2 \\ -0.2}\)
    per cent (\(182\, 000\substack{+5000 \\ -6000}\ \Msun\)),
    largely concentrated within the central regions of the cluster and driven,
    not by a top-heavy IMF producing more BHs initially
    (\(\alpha_3=2.28\substack{+0.08 \\ -0.02}\), close to
    Salpeter; \paperI), but by a very large BH retention fraction
    (\(35\substack{+9 \\ -3}\) per cent).
    This amount of BHs is substantially higher than any other cluster in our
    sample, but is in good agreement with the results of other studies
    \citep[e.g.][]{Zocchi2019,Baumgardt2019b}.

    \citet{Zocchi2019} modelled \omegacen using two-component (one representing
    stellar-mass BHs and one capturing all other lower-mass remnants and
    visible stars) \limepy models.
    Our agreement with their results is interesting, given our inclusion of the
    full mass spectrum, and our fitting of the visible mass function, and
    reinforces the assertion of \citet{Zocchi2019} that a two-component model
    is a reasonable approximation when modelling \omegacen, given its large
    amount of BHs, long two-body relaxation time, and young dynamical age.

    Many claims have been made for the presence of an IMBH in the centre of
    \omegacen. As in the studies of \citet{Zocchi2019} and
    \citet{Baumgardt2019b}, our models suggest that an IMBH is not needed
    to match the data, however we are also limited by the extent of the
    kinematical data available in the very centre of the cluster.
    As was also noted in \citet[][Figure 5]{Zocchi2019}, our models would be
    discrepant with the velocity dispersion profiles of the different
    IMBH-containing models presented by \citet{Noyola2008},
    \citet{vanderMarel2010} and \citet{Baumgardt2017} mostly within the
    innermost few arcseconds of the cluster, where data is currently lacking.
    As such we cannot say for certain
    whether some of the dark mass we find may actually be in the form of a
    central IMBH, given the degeneracy between the effects produced by such an
    IMBH and a central concentration of smaller BHs.

    There is one caveat to our results; the Gaia proper motion
    anisotropy profile shows that
    \omegacen transitions, at about \SI{20}{arcmin} from the centre, from being
    radially anisotropic to being slightly
    tangentially anisotropic. Our
    \limepy models are unable to reproduce any
    amount of tangential anisotropy \citep{Gieles2015}, and thus cannot match
    this feature. Instead, when tangentially biased anisotropy is present in
    our data. the models will favour a mostly isotropic fit as a compromise
    between the radial and tangential regimes \citep{Peuten2017}.
    There is a degeneracy present between the degree of radial
    anisotropy in a cluster and its mass in black holes \citep{Zocchi2017},
    however the difference in the BH mass fraction between the isotropic and
    radially anisotropic models of \citet{Zocchi2019} is only on the order of
    \(\sim 0.7\) per cent. While further exploration of the effects of
    tangential anisotropy on mass models of \omegacen would be interesting,
    given our excellent fits of all other datasets, this should have a
    negligible impact on the results presented here.

\subsubsection{\NGC104}

\begin{figure*}%[h]
        \centering
        \includegraphics[width=\linewidth]{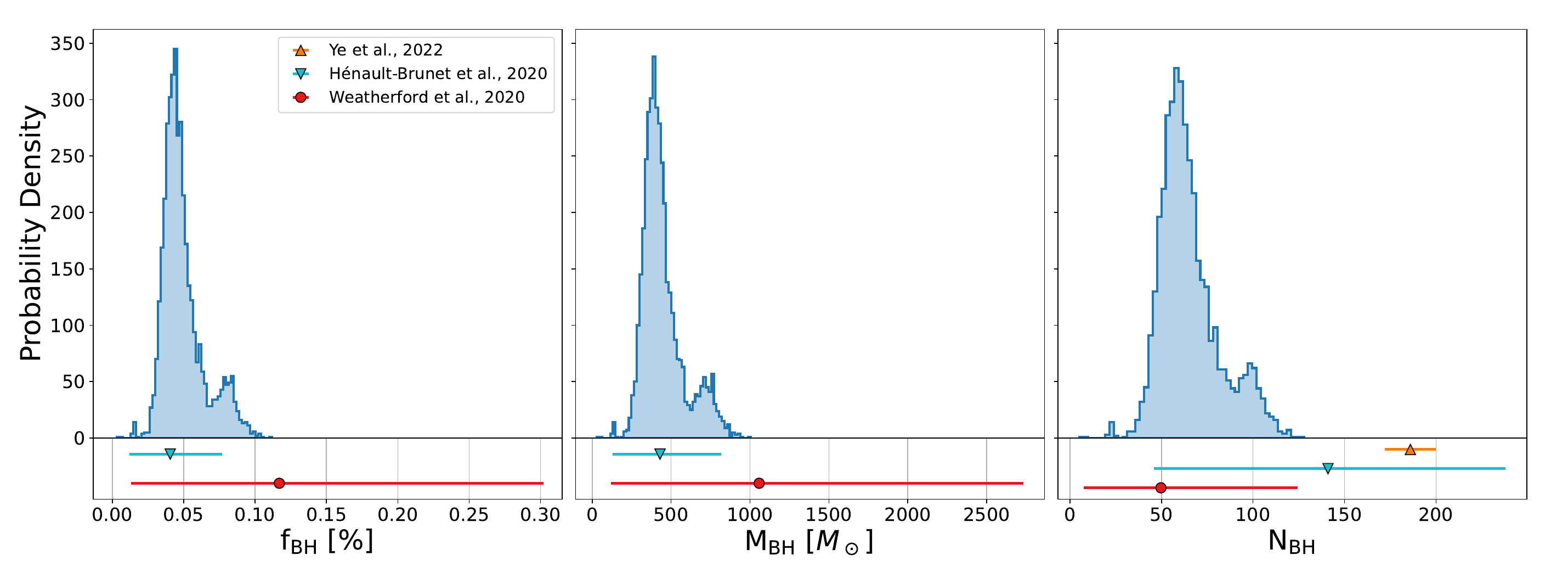}
        \caption{Posterior probability distributions of the BH mass fraction,
                 total mass and number of BHs in 47 Tuc. The results (median and
                 \(1\sigma\)) of various recently inferred values from the
                 literature are shown in the bottom panels.
                 }
        \label{fig:NGC0104_BH_hist}
    \end{figure*}

    \NGC104, or 47 Tuc, is one of the nearest and most massive Milky Way GCs,
    and as such has been extensively studied in the past. Recent modelling
    efforts using Monte Carlo cluster models \citep{Weatherford2020,Ye2022},
    action-integral based DF models \citep{DellaCroce2023}
    and multimass \limepy models \citep{Henault-Brunet2020}
    have provided predictions on the amount of BHs in the cluster,
    while candidate stellar and intermediate mass BHs
    \citep{Millerjones2015,Paduano2024} have been detected within the cluster
    through combined radio and X-ray observations.
    As shown in \Cref{fig:NGC0104_BH_hist}, our models tend to favour a similar
    amount of mass in BHs
    (\(\fbh = 0.046\substack{+0.017 \\ -0.009}\) per cent) to
    other modelling studies, and are consistent within \(2\sigma\)
    with all.
    It is notable that we are able to not only constrain an upper limit on
    the mass in BHs (such as was presented in \citealt{Henault-Brunet2020}),
    but also now place clear and tight bounds on it, thanks to the updated
    treatment of BHs in our multimass models and the updated mass function
    data we are using.

    It was postulated by \citet{Kiziltan2017} that 47 Tuc may host an
    IMBH of around \SI{2300}{\Msun}, based on the analysis of the
    accelerations of millisecond pulsars in the cluster and comparisons with
    \Nbody simulations. However follow-up
    studies using equilibrium models fit to various cluster observables
    (\citealp{Henault-Brunet2020,Mann2019}, although see \citealp{Mann2020})
    determined that there was no need for an IMBH to explain
    the observations, and that a central concentration of less-massive dark
    remnants could explain the data.
    \citet{DellaCroce2023} placed an upper limit on the dark mass
    in the centre of 47 Tuc at \SI{578}{\Msun}, based on the 3D kinematics of
    inner stars.
    These conclusions are again reinforced by our results, which favour
    a smaller central concentration of stellar-mass black holes, alongside a
    population of white dwarfs and neutron stars.

\subsubsection{\NGC6397}

    \NGC6397 is a metal-poor, core-collapsed Milky Way GC at a very short
    heliocentric distance (\(\sim \SI{2.4}{\rm kpc}\);
    \citealp[][2010 edition]{Harris}),
    which has been well studied in the past.
    \citet{Kamann2016} first showed that models including an IMBH or very
    centrally concentrated cluster of stellar-mass BHs of
    \(\sim \SI{600}{\Msun}\) could best reproduce the central kinematics of this
    cluster. \citet{Vitral2021} showed, in turn, that Jeans models with more
    robust proper motion fitting disfavoured an IMBH, and instead proposed an
    inner subcluster of unresolved dark remnants measuring
    \(\sim 1000 \rm{-} 2000 \ \Msun\), which they suggested is dominated by
    stellar-mass BHs.
    However, \citet{Rui2021b,Rui2021a} demonstrated, through fits
    of CMC models, that no BHs were required to explain the kinematics
    of \NGC6397 and that the inner density profile of the cluster argues
    against the presence of BHs. This is in line with the core-collapsed nature
    of \NGC6397 and reinforced by the mass segregation based estimates of
    \citet{Weatherford2020}. These results suggest instead that the central
    dark subcluster could be made up largely of white dwarfs \citep{Kremer2021}.
    A subsequent re-examination of the Jeans modelling of \NGC6397 by
    \citet{Vitral2022}, with updated proper motion datasets, lowered the
    claimed mass of the central ``dark'' cluster to \(\sim \SI{800}{\Msun}\),
    and concurred with a subcluster dominated by white dwarfs, instead of
    stellar-mass BHs.

    Our best-fitting models of \NGC6397, despite
    the caveats of modelling core-collapsed clusters discussed in
    \Cref{sec:core_collapsed_clusters}, favour a negligible population of black
    holes (\(\fbh = 0.019\substack{+0.001 \\ -0.001}\) per cent),
    consistent with the results of \citet{Weatherford2020} and \citet{Rui2021b}.
    Our models also favour a large population of white dwarfs and neutron stars
    dominating the core of the cluster,
    and it is clear that they concur with the general consensus that
    \NGC6397 hosts a massive central concentration of WDs, and little to
    no BHs, nor an IMBH.

\subsubsection{\NGC3201}\label{subsubsec:NGC3201}

    \NGC3201 is a nearby Milky Way GC which has a notably low and flat core
    density profile (i.e. far from core-collapsed), and is the host of three
    confirmed stellar-mass black hole candidates in detached binaries
    \citep{Giesers2018,Giesers2019}.

    CMC models of \NGC3201 \citep{Kremer2018,Kremer2019} suggested that
    models with \(\sim 120\) BHs were best able to
    recreate the velocity dispersion and surface brightness profiles, in
    general agreement with the results of \citet{Askar2018} and the inner
    subcluster of dark remnants found by \citet{Vitral2022}.
    \citet{Weatherford2020} in turn favoured a slightly lower, but still
    consistent, \(\sim 44\) BHs (\(\fbh\sim0.6\) per cent),
    with large uncertainties.
    In contrast, our best-fitting models of \NGC3201 favour a remarkably small
    amount of BHs, with the distribution peaking at 0 BH
    (95\% probability of containing less than 10 BHs,
    \(\fbh = 0.007\substack{+0.02 \\ -0.007}\) per cent).
    This is somewhat surprising, given the literature results and the shape of
    the cluster density profile, but is technically in agreement (within
    errors) with the results from the literature,
    and follows the trend in our results of predicting fewer BHs than other
    studies in this regime (see \Cref{fig:f_BH_rescomp}).

    As was noted in \Cref{sec:validation}, \NGC3201 stands out in our sample as
    a cluster with a long relaxation time and young dynamical age which was
    best fit by a high \(\delta\) value (and thus low number of BHs).
    As discussed above, this is unexpected for dynamically young
    clusters, and is likely indicative that this cluster
    falls within a regime our models may struggle to capture.
    Therefore the results for this cluster based on our multimass model
    fit should be taken with extra caution.
    \NGC3201 is the cluster in our sample for which this issue is
    potentially the most severe.

\subsubsection{\NGC6121}

    \NGC6121, or \Messier4, is the nearest GC (\(d \sim \SI{1.85}{\kilo\pc}\);
    \citealp{Baumgardt2021}) and as such has been extensively observed.
    Recently, \citet{Vitral2023} utilized Jeans modelling of \Messier4,
    followed up by  a comparison with Monte Carlo models, to fit on \HST and
    \Gaia kinematic data and suggest an excess of dark mass
    of around \(800 \pm \SI{300}{\Msun}\), concentrated within the inner
    0.016-\SI{0.034}{\parsec} of the cluster.
    They also explore the possibility that this mass is concentrated in a single
    IMBH.
    Our models have a comparable amount of dark mass within the central regions
    (dominated almost entirely by white dwarfs) but in a less concentrated
    mass profile, reaching a similar cumulative mass of around \SI{800}{\Msun}
    near \SI{0.1}{\parsec} instead. Our models also suggest a small population
    of BHs, totalling around \(80\substack{+50 \\ -30}\ \Msun\)
    (\(\fbh=0.09\substack{+0.05 \\ -0.03}\) per cent).
    This mass in BHs is significantly smaller than the dark mass suggested by
    the Jeans models of \citet{Vitral2023}, however, notably, this is actually
    in good agreement with the best matching CMC model they also presented.

    One key difference between our analyses lies in the data used.
    \citet{Vitral2023} introduce new HST proper motion data which reaches
    deeper into the core of the cluster than the \citet{Libralato2022} data
    we use and their velocity dispersion profile seems to increase toward the
    centre in the inner few arcseconds, albeit with very large uncertainties.
    However, given these large uncertainties and our very good fits to all
    other datasets, it is unlikely that our results would
    change significantly with the inclusion of these new datapoints.

%---------------------------------------------------------------------------
% Conclusions
%---------------------------------------------------------------------------

% !TEX root = ./paper.tex

%---------------------------------------------------------------------------
\section{Conclusions}\label{sec:conclusions}
%---------------------------------------------------------------------------

    % what we did

    In this work, we have utilized the best-fitting multimass models of 34
    GCs, first presented in \paperI, to explore the BH and remnant
    populations of a large sample of Milky Way clusters, yielding a number of
    important conclusions:

    \begin{enumerate}

    % summarize bh results

        \item
        The models allow us to infer best-fitting, posterior probability
        distributions for the total mass, number and mass fraction of BHs in
        our sample of clusters.
        These results indicate that a large number of the GCs are
        consistent with hosting little to no BHs, with the largest BH
        populations reaching masses in BHs up to a few thousand \Msun and
        mass fractions of around 1 per cent (save for \omegacen,
        for which \(\fbh \sim 5\) per cent).

        \item
        We find an anti-correlation between the BH mass fraction
        and the \(\delta\) parameter, a proxy of mass segregation, with clusters
        having little mass in BHs congregating around \(\sim 0.5\), while
        \(\delta\) is lower (closer to 0.3) and mass
        segregation is increasingly suppressed in clusters with more substantial
        BH populations, in agreement with the findings of \citet{Peuten2017}.

        \item
        As the \fbh we see in the clusters of our sample are well below the
        critical mass fraction of \(\sim2.5\) per cent noted by
        \citet{Gieles2021}, these clusters are expected to continue to evolve
        towards a point when they will have ejected all their BHs.
        The inferred present-day \fbh encode information about the dynamical
        evolution and initial density of GCs that can be used in future work
        to infer the initial conditions of the population of Milky Way GCs.

        \item
        A clear correlation is also found between the dynamical age of the
        clusters and the overall remnant mass fraction,
        which increases as clusters evolve
        and lose low-mass stars. Our results show that the most evolved GCs in
        our sample are made up of around 70 per cent dark remnants, by mass,
        at the present day.

        \item
        We find typically good agreement overall, within uncertainties, between
        our results and those of other studies inferring BH populations in GCs
        in the literature \cite[e.g.][]{Askar2018,Weatherford2020}, but with
        notable discrepancies between individual clusters.
        Our inferred masses in BHs are, generally, slightly smaller (larger)
        than these studies in clusters with small (large) BH populations.

        \item
        Closer inspection of a number of interesting clusters with previous
        claims of hosting an elusive IMBH reveal no need for
        such an object to explain the large amount of data used in our model
        fitting.

    \end{enumerate}

%---------------------------------------------------------------------------
\section*{Acknowledgements}
%---------------------------------------------------------------------------

ND is grateful for the support of the Durland Scholarship in Graduate Research.
VHB acknowledges the support of the Natural Sciences and Engineering Research
Council of Canada (NSERC) through grant RGPIN-2020-05990.
MG acknowledges support from the Ministry of Science and Innovation
(EUR2020-112157, PID2021-125485NB-C22, CEX2019-000918-M funded by
MCIN/AEI/10.13039/501100011033) and from AGAUR (SGR-2021-01069).

This research was enabled in part by support provided by ACENET
(\url{www.ace-net.ca}) and the Digital Research Alliance of Canada
(\url{https://alliancecan.ca}).

This work has also benefited from a variety of \texttt{Python}
packages including
\texttt{astropy} \citep{Astropy2013,Astropy2018},
\texttt{dynesty} \citep{Speagle2020},
\texttt{emcee} \citep{Foreman-Mackey2016},
\texttt{h5py} \citep{Collette2022},
\texttt{cmctoolkit} \citep{cmctoolkit},
\texttt{JAX} \citep{jax2018github},
\texttt{blackjax} \citep{blackjax2020github},
\texttt{matplotlib} \citep{Hunter2007},
\texttt{numpy} \citep{Harris2020},
\texttt{scipy} \citep{Virtanen2020} and
\texttt{shapely} \citep{Gillies2022}.

%---------------------------------------------------------------------------
\section*{Data Availability}
%---------------------------------------------------------------------------

The data underlying this article are available at
\url{https://github.com/nmdickson/GCfit-results}.

%---------------------------------------------------------------------------
% REFERENCES
%---------------------------------------------------------------------------

\bibliographystyle{mnras}
\bibliography{biblio}

\begin{thebibliography}{}
\makeatletter
\relax
\def\mn@urlcharsother{\let\do\@makeother \do\$\do\&\do\#\do\^\do\_\do\%\do\~}
\def\mn@doi{\begingroup\mn@urlcharsother \@ifnextchar [ {\mn@doi@}
  {\mn@doi@[]}}
\def\mn@doi@[#1]#2{\def\@tempa{#1}\ifx\@tempa\@empty \href
  {http://dx.doi.org/#2} {doi:#2}\else \href {http://dx.doi.org/#2} {#1}\fi
  \endgroup}
\def\mn@eprint#1#2{\mn@eprint@#1:#2::\@nil}
\def\mn@eprint@arXiv#1{\href {http://arxiv.org/abs/#1} {{\tt arXiv:#1}}}
\def\mn@eprint@dblp#1{\href {http://dblp.uni-trier.de/rec/bibtex/#1.xml}
  {dblp:#1}}
\def\mn@eprint@#1:#2:#3:#4\@nil{\def\@tempa {#1}\def\@tempb {#2}\def\@tempc
  {#3}\ifx \@tempc \@empty \let \@tempc \@tempb \let \@tempb \@tempa \fi \ifx
  \@tempb \@empty \def\@tempb {arXiv}\fi \@ifundefined
  {mn@eprint@\@tempb}{\@tempb:\@tempc}{\expandafter \expandafter \csname
  mn@eprint@\@tempb\endcsname \expandafter{\@tempc}}}

\bibitem[\protect\citeauthoryear{{Abbate}, {Possenti}, {Colpi}  \&
  {Spera}}{{Abbate} et~al.}{2019}]{Abbate2019}
{Abbate} F.,  {Possenti} A.,  {Colpi} M.,   {Spera} M.,  2019, \mn@doi [\apjl]
  {10.3847/2041-8213/ab46c3}, 884, L9

\bibitem[\protect\citeauthoryear{{Abbott} et~al.,}{{Abbott}
  et~al.}{2016}]{Abbott2016a}
{Abbott} B.~P.,  et~al., 2016, \mn@doi [\apjl] {10.3847/2041-8205/818/2/L22},
  \href {http://adsabs.harvard.edu/abs/2016ApJ...818L..22A} {818, L22}

\bibitem[\protect\citeauthoryear{{Alessandrini}, {Lanzoni}, {Ferraro},
  {Miocchi}  \& {Vesperini}}{{Alessandrini} et~al.}{2016}]{Alessandrini2016}
{Alessandrini} E.,  {Lanzoni} B.,  {Ferraro} F.~R.,  {Miocchi} P.,
  {Vesperini} E.,  2016, \mn@doi [\apj] {10.3847/1538-4357/833/2/252}, 833, 252

\bibitem[\protect\citeauthoryear{{Antonini} \& {Gieles}}{{Antonini} \&
  {Gieles}}{2020a}]{Antonini2020b}
{Antonini} F.,  {Gieles} M.,  2020a, \mn@doi [\prd]
  {10.1103/PhysRevD.102.123016}, 102, 123016

\bibitem[\protect\citeauthoryear{{Antonini} \& {Gieles}}{{Antonini} \&
  {Gieles}}{2020b}]{Antonini2020a}
{Antonini} F.,  {Gieles} M.,  2020b, \mn@doi [\mnras] {10.1093/mnras/stz3584},
  492, 2936

\bibitem[\protect\citeauthoryear{{Antonini}, {Gieles}, {Dosopoulou}  \&
  {Chattopadhyay}}{{Antonini} et~al.}{2023}]{Antonini2023}
{Antonini} F.,  {Gieles} M.,  {Dosopoulou} F.,   {Chattopadhyay} D.,  2023,
  \mn@doi [\mnras] {10.1093/mnras/stad972}, 522, 466

\bibitem[\protect\citeauthoryear{{Arca Sedda}, {Askar}  \& {Giersz}}{{Arca
  Sedda} et~al.}{2018}]{ArcaSedda2018}
{Arca Sedda} M.,  {Askar} A.,   {Giersz} M.,  2018, \mn@doi [\mnras]
  {10.1093/mnras/sty1859}, 479, 4652

\bibitem[\protect\citeauthoryear{{Aros} \& {Vesperini}}{{Aros} \&
  {Vesperini}}{2023}]{Aros2023}
{Aros} F.~I.,  {Vesperini} E.,  2023, \mn@doi [arXiv e-prints]
  {10.48550/arXiv.2308.03845}, p. arXiv:2308.03845

\bibitem[\protect\citeauthoryear{{Askar}, {Arca Sedda}  \& {Giersz}}{{Askar}
  et~al.}{2018}]{Askar2018}
{Askar} A.,  {Arca Sedda} M.,   {Giersz} M.,  2018, \mn@doi [\mnras]
  {10.1093/mnras/sty1186}, 478, 1844

\bibitem[\protect\citeauthoryear{{Astropy Collaboration} et~al.,}{{Astropy
  Collaboration} et~al.}{2013}]{Astropy2013}
{Astropy Collaboration} et~al., 2013, \mn@doi [\aap]
  {10.1051/0004-6361/201322068}, 558, A33

\bibitem[\protect\citeauthoryear{{Astropy Collaboration} et~al.,}{{Astropy
  Collaboration} et~al.}{2018}]{Astropy2018}
{Astropy Collaboration} et~al., 2018, \mn@doi [\aj] {10.3847/1538-3881/aabc4f},
  156, 123

\bibitem[\protect\citeauthoryear{{Balbinot} \& {Gieles}}{{Balbinot} \&
  {Gieles}}{2018}]{Balbinot2018}
{Balbinot} E.,  {Gieles} M.,  2018, \mn@doi [\mnras] {10.1093/mnras/stx2708},
  474, 2479

\bibitem[\protect\citeauthoryear{{Banerjee} \& {Kroupa}}{{Banerjee} \&
  {Kroupa}}{2011}]{Banerjee2011}
{Banerjee} S.,  {Kroupa} P.,  2011, \mn@doi [\apjl]
  {10.1088/2041-8205/741/1/L12}, 741, L12

\bibitem[\protect\citeauthoryear{{Banerjee}, {Belczynski}, {Fryer}, {Berczik},
  {Hurley}, {Spurzem}  \& {Wang}}{{Banerjee} et~al.}{2020}]{Banerjee2020}
{Banerjee} S.,  {Belczynski} K.,  {Fryer} C.~L.,  {Berczik} P.,  {Hurley}
  J.~R.,  {Spurzem} R.,   {Wang} L.,  2020, \mn@doi [\aap]
  {10.1051/0004-6361/201935332}, 639, A41

\bibitem[\protect\citeauthoryear{{Baumgardt}}{{Baumgardt}}{2017}]{Baumgardt2017}
{Baumgardt} H.,  2017, \mn@doi [\mnras] {10.1093/mnras/stw2488}, 464, 2174

\bibitem[\protect\citeauthoryear{{Baumgardt} \& {Hilker}}{{Baumgardt} \&
  {Hilker}}{2018}]{Baumgardt2018}
{Baumgardt} H.,  {Hilker} M.,  2018, \mn@doi [\mnras] {10.1093/mnras/sty1057},
  478, 1520

\bibitem[\protect\citeauthoryear{{Baumgardt} \& {Makino}}{{Baumgardt} \&
  {Makino}}{2003}]{Baumgardt2003}
{Baumgardt} H.,  {Makino} J.,  2003, \mn@doi [\mnras]
  {10.1046/j.1365-8711.2003.06286.x}, 340, 227

\bibitem[\protect\citeauthoryear{{Baumgardt} \& {Vasiliev}}{{Baumgardt} \&
  {Vasiliev}}{2021}]{Baumgardt2021}
{Baumgardt} H.,  {Vasiliev} E.,  2021, \mn@doi [\mnras]
  {10.1093/mnras/stab1474}, 505, 5957

\bibitem[\protect\citeauthoryear{{Baumgardt}, {Parmentier}, {Gieles}  \&
  {Vesperini}}{{Baumgardt} et~al.}{2010}]{Baumgardt2010}
{Baumgardt} H.,  {Parmentier} G.,  {Gieles} M.,   {Vesperini} E.,  2010,
  \mn@doi [\mnras] {10.1111/j.1365-2966.2009.15758.x}, 401, 1832

\bibitem[\protect\citeauthoryear{{Baumgardt}, {Hilker}, {Sollima}  \&
  {Bellini}}{{Baumgardt} et~al.}{2019a}]{Baumgardt2019a}
{Baumgardt} H.,  {Hilker} M.,  {Sollima} A.,   {Bellini} A.,  2019a, \mn@doi
  [\mnras] {10.1093/mnras/sty2997}, 482, 5138

\bibitem[\protect\citeauthoryear{{Baumgardt} et~al.,}{{Baumgardt}
  et~al.}{2019b}]{Baumgardt2019b}
{Baumgardt} H.,  et~al., 2019b, \mn@doi [\mnras] {10.1093/mnras/stz2060}, 488,
  5340

\bibitem[\protect\citeauthoryear{{Baumgardt}, {Sollima}  \&
  {Hilker}}{{Baumgardt} et~al.}{2020}]{Baumgardt2020}
{Baumgardt} H.,  {Sollima} A.,   {Hilker} M.,  2020, \mn@doi [\pasa]
  {10.1017/pasa.2020.38}, 37, e046

\bibitem[\protect\citeauthoryear{Baumgardt, Hénault-Brunet, Dickson  \&
  Sollima}{Baumgardt et~al.}{2023}]{Baumgardt2023}
Baumgardt H.,  Hénault-Brunet V.,  Dickson N.,   Sollima A.,  2023, \mn@doi
  [\mnras] {10.1093/mnras/stad631}, 521, 3991

\bibitem[\protect\citeauthoryear{Bradbury et~al.,}{Bradbury
  et~al.}{2018}]{jax2018github}
Bradbury J.,  et~al., 2018, {JAX}: composable transformations of
  {P}ython+{N}um{P}y programs, \url {http://github.com/google/jax}

\bibitem[\protect\citeauthoryear{{Breen} \& {Heggie}}{{Breen} \&
  {Heggie}}{2013a}]{Breen2013a}
{Breen} P.~G.,  {Heggie} D.~C.,  2013a, \mn@doi [\mnras]
  {10.1093/mnras/stt628}, 432, 2779

\bibitem[\protect\citeauthoryear{{Breen} \& {Heggie}}{{Breen} \&
  {Heggie}}{2013b}]{Breen2013b}
{Breen} P.~G.,  {Heggie} D.~C.,  2013b, \mn@doi [\mnras]
  {10.1093/mnras/stt1599}, 436, 584

\bibitem[\protect\citeauthoryear{{Chatterjee}, {Umbreit}, {Fregeau}  \&
  {Rasio}}{{Chatterjee} et~al.}{2013}]{Chatterjee2013}
{Chatterjee} S.,  {Umbreit} S.,  {Fregeau} J.~M.,   {Rasio} F.~A.,  2013,
  \mn@doi [\mnras] {10.1093/mnras/sts464}, 429, 2881

\bibitem[\protect\citeauthoryear{{Chatterjee}, {Rodriguez}  \&
  {Rasio}}{{Chatterjee} et~al.}{2017}]{Chatterjee2017}
{Chatterjee} S.,  {Rodriguez} C.~L.,   {Rasio} F.~A.,  2017, \mn@doi [\apj]
  {10.3847/1538-4357/834/1/68}, 834, 68

\bibitem[\protect\citeauthoryear{{Choi}, {Dotter}, {Conroy}, {Cantiello},
  {Paxton}  \& {Johnson}}{{Choi} et~al.}{2016}]{Choi2016}
{Choi} J.,  {Dotter} A.,  {Conroy} C.,  {Cantiello} M.,  {Paxton} B.,
  {Johnson} B.~D.,  2016, \mn@doi [\apj] {10.3847/0004-637X/823/2/102}, 823,
  102

\bibitem[\protect\citeauthoryear{Collette et~al.,}{Collette
  et~al.}{2022}]{Collette2022}
Collette A.,  et~al., 2022, h5py: 3.7.0, \mn@doi{10.5281/zenodo.6575970}

\bibitem[\protect\citeauthoryear{{Dalgleish} et~al.,}{{Dalgleish}
  et~al.}{2020}]{Dalgleish2020}
{Dalgleish} H.,  et~al., 2020, \mn@doi [\mnras] {10.1093/mnras/staa091}, 492,
  3859

\bibitem[\protect\citeauthoryear{{Della Croce}, {Pascale}, {Giunchi}, {Nipoti},
  {Cignoni}  \& {Dalessandro}}{{Della Croce} et~al.}{2023}]{DellaCroce2023}
{Della Croce} A.,  {Pascale} R.,  {Giunchi} E.,  {Nipoti} C.,  {Cignoni} M.,
  {Dalessandro} E.,  2023, \mn@doi [arXiv e-prints]
  {10.48550/arXiv.2310.15221}, p. arXiv:2310.15221

\bibitem[\protect\citeauthoryear{{Dickson}, {H{\'e}nault-Brunet}, {Baumgardt},
  {Gieles}  \& {Smith}}{{Dickson} et~al.}{2023}]{Dickson2023}
{Dickson} N.,  {H{\'e}nault-Brunet} V.,  {Baumgardt} H.,  {Gieles} M.,
  {Smith} P.~J.,  2023, \mn@doi [\mnras] {10.1093/mnras/stad1254}, 522, 5320

\bibitem[\protect\citeauthoryear{{Djorgovski} \& {King}}{{Djorgovski} \&
  {King}}{1986}]{Djorgovski1986}
{Djorgovski} S.,  {King} I.~R.,  1986, \apjl, 305, L61

\bibitem[\protect\citeauthoryear{{Dotter}}{{Dotter}}{2016}]{Dotter2016}
{Dotter} A.,  2016, \mn@doi [\apjs] {10.3847/0067-0049/222/1/8}, 222, 8

\bibitem[\protect\citeauthoryear{{Dotter}, {Chaboyer}, {Jevremovi{\'c}},
  {Baron}, {Ferguson}, {Sarajedini}  \& {Anderson}}{{Dotter}
  et~al.}{2007}]{Dotter2007}
{Dotter} A.,  {Chaboyer} B.,  {Jevremovi{\'c}} D.,  {Baron} E.,  {Ferguson}
  J.~W.,  {Sarajedini} A.,   {Anderson} J.,  2007, \mn@doi [\aj]
  {10.1086/517915}, 134, 376

\bibitem[\protect\citeauthoryear{{Dotter}, {Chaboyer}, {Jevremovi{\'c}},
  {Kostov}, {Baron}  \& {Ferguson}}{{Dotter} et~al.}{2008}]{Dotter2008}
{Dotter} A.,  {Chaboyer} B.,  {Jevremovi{\'c}} D.,  {Kostov} V.,  {Baron} E.,
  {Ferguson} J.~W.,  2008, \mn@doi [\apjs] {10.1086/589654}, 178, 89

\bibitem[\protect\citeauthoryear{{Foreman-Mackey}}{{Foreman-Mackey}}{2016}]{Foreman-Mackey2016}
{Foreman-Mackey} D.,  2016, \mn@doi [The Journal of Open Source Software]
  {10.21105/joss.00024}, 1, 24

\bibitem[\protect\citeauthoryear{{Fryer}, {Belczynski}, {Wiktorowicz},
  {Dominik}, {Kalogera}  \& {Holz}}{{Fryer} et~al.}{2012}]{Fryer2012}
{Fryer} C.~L.,  {Belczynski} K.,  {Wiktorowicz} G.,  {Dominik} M.,  {Kalogera}
  V.,   {Holz} D.~E.,  2012, \mn@doi [\apj] {10.1088/0004-637X/749/1/91}, 749,
  91

\bibitem[\protect\citeauthoryear{{Gaia Collaboration} et~al.,}{{Gaia
  Collaboration} et~al.}{2022}]{GaiaCollaboration2022}
{Gaia Collaboration} et~al., 2022, \mn@doi [arXiv e-prints]
  {10.48550/arXiv.2208.00211}, p. arXiv:2208.00211

\bibitem[\protect\citeauthoryear{{Gieles} \& {Gnedin}}{{Gieles} \&
  {Gnedin}}{2023}]{Gieles2023}
{Gieles} M.,  {Gnedin} O.~Y.,  2023, \mn@doi [\mnras] {10.1093/mnras/stad1287},
  522, 5340

\bibitem[\protect\citeauthoryear{{Gieles} \& {Zocchi}}{{Gieles} \&
  {Zocchi}}{2015}]{Gieles2015}
{Gieles} M.,  {Zocchi} A.,  2015, \mn@doi [mnras] {10.1093/mnras/stv1848}, 454,
  576

\bibitem[\protect\citeauthoryear{{Gieles}, {Heggie}  \& {Zhao}}{{Gieles}
  et~al.}{2011}]{Gieles2011}
{Gieles} M.,  {Heggie} D.~C.,   {Zhao} H.,  2011, \mn@doi [\mnras]
  {10.1111/j.1365-2966.2011.18320.x}, 413, 2509

\bibitem[\protect\citeauthoryear{{Gieles}, {Balbinot}, {Yaaqib},
  {H{\'e}nault-Brunet}, {Zocchi}, {Peuten}  \& {Jonker}}{{Gieles}
  et~al.}{2018}]{Gieles2018}
{Gieles} M.,  {Balbinot} E.,  {Yaaqib} R. I.~S.~M.,  {H{\'e}nault-Brunet} V.,
  {Zocchi} A.,  {Peuten} M.,   {Jonker} P.~G.,  2018, \mn@doi [\mnras]
  {10.1093/mnras/stx2694}, 473, 4832

\bibitem[\protect\citeauthoryear{{Gieles}, {Erkal}, {Antonini}, {Balbinot}  \&
  {Pe{\~n}arrubia}}{{Gieles} et~al.}{2021}]{Gieles2021}
{Gieles} M.,  {Erkal} D.,  {Antonini} F.,  {Balbinot} E.,   {Pe{\~n}arrubia}
  J.,  2021, \mn@doi [Nature Astronomy] {10.1038/s41550-021-01392-2}, 5, 957

\bibitem[\protect\citeauthoryear{{Giesers} et~al.,}{{Giesers}
  et~al.}{2018}]{Giesers2018}
{Giesers} B.,  et~al., 2018, \mn@doi [\mnras] {10.1093/mnrasl/slx203}, 475, L15

\bibitem[\protect\citeauthoryear{{Giesers} et~al.,}{{Giesers}
  et~al.}{2019}]{Giesers2019}
{Giesers} B.,  et~al., 2019, \mn@doi [\aap] {10.1051/0004-6361/201936203}, 632,
  A3

\bibitem[\protect\citeauthoryear{Gillies, van~der Wel, Van~den Bossche, Taves,
  Arnott, Ward  et~al.}{Gillies et~al.}{2022}]{Gillies2022}
Gillies S.,  van~der Wel C.,  Van~den Bossche J.,  Taves M.~W.,  Arnott J.,
  Ward B.~C.,   et~al., 2022, Shapely, \mn@doi{10.5281/zenodo.7263102}

\bibitem[\protect\citeauthoryear{{Harris}}{{Harris}}{1996}]{Harris}
{Harris} W.~E.,  1996, \mn@doi [\aj] {10.1086/118116}, 112, 1487

\bibitem[\protect\citeauthoryear{{Harris} et~al.,}{{Harris}
  et~al.}{2020}]{Harris2020}
{Harris} C.~R.,  et~al., 2020, \mn@doi [\nat] {10.1038/s41586-020-2649-2}, 585,
  357

\bibitem[\protect\citeauthoryear{{H{\'e}nault-Brunet}, {Gieles}, {Sollima},
  {Watkins}, {Zocchi}, {Claydon}, {Pancino}  \&
  {Baumgardt}}{{H{\'e}nault-Brunet} et~al.}{2019}]{Henault-Brunet2019}
{H{\'e}nault-Brunet} V.,  {Gieles} M.,  {Sollima} A.,  {Watkins} L.~L.,
  {Zocchi} A.,  {Claydon} I.,  {Pancino} E.,   {Baumgardt} H.,  2019, \mn@doi
  [mnras] {10.1093/mnras/sty3187}, 483, 1400

\bibitem[\protect\citeauthoryear{{H{\'e}nault-Brunet}, {Gieles}, {Strader},
  {Peuten}, {Balbinot}  \& {Douglas}}{{H{\'e}nault-Brunet}
  et~al.}{2020}]{Henault-Brunet2020}
{H{\'e}nault-Brunet} V.,  {Gieles} M.,  {Strader} J.,  {Peuten} M.,  {Balbinot}
  E.,   {Douglas} K.~E.~K.,  2020, \mn@doi [\mnras] {10.1093/mnras/stz2995},
  491, 113

\bibitem[\protect\citeauthoryear{{H{\'e}non}}{{H{\'e}non}}{1961}]{Henon1961}
{H{\'e}non} M.,  1961, Ann. Astrophys., 24, 369; English trans.:
  arXiv:1103.3499

\bibitem[\protect\citeauthoryear{{Hobbs}, {Lorimer}, {Lyne}  \&
  {Kramer}}{{Hobbs} et~al.}{2005}]{Hobbs2005}
{Hobbs} G.,  {Lorimer} D.~R.,  {Lyne} A.~G.,   {Kramer} M.,  2005, \mn@doi
  [\mnras] {10.1111/j.1365-2966.2005.09087.x}, 360, 974

\bibitem[\protect\citeauthoryear{{Hunter}}{{Hunter}}{2007}]{Hunter2007}
{Hunter} J.~D.,  2007, \mn@doi [Computing in Science and Engineering]
  {10.1109/MCSE.2007.55}, 9, 90

\bibitem[\protect\citeauthoryear{{Hurley}}{{Hurley}}{2007}]{Hurley2007}
{Hurley} J.~R.,  2007, \mn@doi [\mnras] {10.1111/j.1365-2966.2007.11912.x},
  379, 93

\bibitem[\protect\citeauthoryear{{Hurley}, {Sippel}, {Tout}  \&
  {Aarseth}}{{Hurley} et~al.}{2016}]{Hurley2016}
{Hurley} J.~R.,  {Sippel} A.~C.,  {Tout} C.~A.,   {Aarseth} S.~J.,  2016,
  \mn@doi [\pasa] {10.1017/pasa.2016.30}, 33, e036

\bibitem[\protect\citeauthoryear{{Kamann}, {Wisotzki}, {Roth}, {Gerssen},
  {Husser}, {Sandin}  \& {Weilbacher}}{{Kamann} et~al.}{2014}]{Kamann2014}
{Kamann} S.,  {Wisotzki} L.,  {Roth} M.~M.,  {Gerssen} J.,  {Husser} T.~O.,
  {Sandin} C.,   {Weilbacher} P.,  2014, \mn@doi [\aap]
  {10.1051/0004-6361/201322183}, 566, A58

\bibitem[\protect\citeauthoryear{{Kamann} et~al.,}{{Kamann}
  et~al.}{2016}]{Kamann2016}
{Kamann} S.,  et~al., 2016, \mn@doi [\aap] {10.1051/0004-6361/201527065}, 588,
  A149

\bibitem[\protect\citeauthoryear{{Kamann} et~al.,}{{Kamann}
  et~al.}{2018}]{Kamann2018}
{Kamann} S.,  et~al., 2018, \mn@doi [\mnras] {10.1093/mnras/stx2719}, 473, 5591

\bibitem[\protect\citeauthoryear{{K{\i}z{\i}ltan}, {Baumgardt}  \&
  {Loeb}}{{K{\i}z{\i}ltan} et~al.}{2017}]{Kiziltan2017}
{K{\i}z{\i}ltan} B.,  {Baumgardt} H.,   {Loeb} A.,  2017, \mn@doi [\nat]
  {10.1038/nature21361}, 542, 203

\bibitem[\protect\citeauthoryear{{Kremer}, {Ye}, {Chatterjee}, {Rodriguez}  \&
  {Rasio}}{{Kremer} et~al.}{2018}]{Kremer2018}
{Kremer} K.,  {Ye} C.~S.,  {Chatterjee} S.,  {Rodriguez} C.~L.,   {Rasio}
  F.~A.,  2018, \mn@doi [\apjl] {10.3847/2041-8213/aab26c}, 855, L15

\bibitem[\protect\citeauthoryear{{Kremer}, {Chatterjee}, {Ye}, {Rodriguez}  \&
  {Rasio}}{{Kremer} et~al.}{2019}]{Kremer2019}
{Kremer} K.,  {Chatterjee} S.,  {Ye} C.~S.,  {Rodriguez} C.~L.,   {Rasio}
  F.~A.,  2019, \mn@doi [\apj] {10.3847/1538-4357/aaf646}, 871, 38

\bibitem[\protect\citeauthoryear{{Kremer} et~al.,}{{Kremer}
  et~al.}{2020a}]{Kremer2020b}
{Kremer} K.,  et~al., 2020a, \mn@doi [\apjs] {10.3847/1538-4365/ab7919}, 247,
  48

\bibitem[\protect\citeauthoryear{{Kremer}, {Ye}, {Chatterjee}, {Rodriguez}  \&
  {Rasio}}{{Kremer} et~al.}{2020b}]{Kremer2020a}
{Kremer} K.,  {Ye} C.~S.,  {Chatterjee} S.,  {Rodriguez} C.~L.,   {Rasio}
  F.~A.,  2020b, in {Bragaglia} A.,  {Davies} M.,  {Sills} A.,   {Vesperini}
  E.,  eds, ~ Vol. 351, Star Clusters: From the Milky Way to the Early
  Universe. pp 357--366, \mn@doi{10.1017/S1743921319007269}

\bibitem[\protect\citeauthoryear{{Kremer}, {Rui}, {Weatherford}, {Chatterjee},
  {Fragione}, {Rasio}, {Rodriguez}  \& {Ye}}{{Kremer}
  et~al.}{2021}]{Kremer2021}
{Kremer} K.,  {Rui} N.~Z.,  {Weatherford} N.~C.,  {Chatterjee} S.,  {Fragione}
  G.,  {Rasio} F.~A.,  {Rodriguez} C.~L.,   {Ye} C.~S.,  2021, \mn@doi [\apj]
  {10.3847/1538-4357/ac06d4}, 917, 28

\bibitem[\protect\citeauthoryear{{Kroupa}}{{Kroupa}}{2001}]{Kroupa2001}
{Kroupa} P.,  2001, \mn@doi [\mnras] {10.1046/j.1365-8711.2001.04022.x}, 322,
  231

\bibitem[\protect\citeauthoryear{Lao \& Louf}{Lao \&
  Louf}{2020}]{blackjax2020github}
Lao J.,  Louf R.,  2020, {B}lackjax: A sampling library for {JAX}, \url
  {http://github.com/blackjax-devs/blackjax}

\bibitem[\protect\citeauthoryear{{Libralato} et~al.,}{{Libralato}
  et~al.}{2022}]{Libralato2022}
{Libralato} M.,  et~al., 2022, \mn@doi [\apj] {10.3847/1538-4357/ac7727}, 934,
  150

\bibitem[\protect\citeauthoryear{{Lindegren} et~al.,}{{Lindegren}
  et~al.}{2021}]{Lindegren2021}
{Lindegren} L.,  et~al., 2021, \mn@doi [\aap] {10.1051/0004-6361/202039709},
  649, A2

\bibitem[\protect\citeauthoryear{{L{\"u}tzgendorf} et~al.,}{{L{\"u}tzgendorf}
  et~al.}{2013}]{Lutzgendorf2013}
{L{\"u}tzgendorf} N.,  et~al., 2013, \mn@doi [\aap]
  {10.1051/0004-6361/201220307}, 552, A49

\bibitem[\protect\citeauthoryear{{Mann} et~al.,}{{Mann}
  et~al.}{2019}]{Mann2019}
{Mann} C.~R.,  et~al., 2019, \mn@doi [\apj] {10.3847/1538-4357/ab0e6d}, 875, 1

\bibitem[\protect\citeauthoryear{{Mann} et~al.,}{{Mann}
  et~al.}{2020}]{Mann2020}
{Mann} C.~R.,  et~al., 2020, \mn@doi [\apj] {10.3847/1538-4357/ab84ea}, 893, 86

\bibitem[\protect\citeauthoryear{{McNamara}, {Harrison}, {Baumgardt}  \&
  {Khalaj}}{{McNamara} et~al.}{2012}]{McNamara2012}
{McNamara} B.~J.,  {Harrison} T.~E.,  {Baumgardt} H.,   {Khalaj} P.,  2012,
  \mn@doi [\apj] {10.1088/0004-637X/745/2/175}, 745, 175

\bibitem[\protect\citeauthoryear{{Meza}, {Navarro}, {Abadi}  \&
  {Steinmetz}}{{Meza} et~al.}{2005}]{Meza2005}
{Meza} A.,  {Navarro} J.~F.,  {Abadi} M.~G.,   {Steinmetz} M.,  2005, \mn@doi
  [\mnras] {10.1111/j.1365-2966.2005.08869.x}, 359, 93

\bibitem[\protect\citeauthoryear{{Miller-Jones} et~al.,}{{Miller-Jones}
  et~al.}{2015}]{Millerjones2015}
{Miller-Jones} J.~C.~A.,  et~al., 2015, \mn@doi [\mnras]
  {10.1093/mnras/stv1869}, 453, 3918

\bibitem[\protect\citeauthoryear{{Miocchi} et~al.,}{{Miocchi}
  et~al.}{2013}]{Miocchi2013}
{Miocchi} P.,  et~al., 2013, \mn@doi [\apj] {10.1088/0004-637X/774/2/151}, 774,
  151

\bibitem[\protect\citeauthoryear{{Morscher}, {Umbreit}, {Farr}  \&
  {Rasio}}{{Morscher} et~al.}{2013}]{Morscher2013}
{Morscher} M.,  {Umbreit} S.,  {Farr} W.~M.,   {Rasio} F.~A.,  2013, \mn@doi
  [\apjl] {10.1088/2041-8205/763/1/L15}, \href
  {https://ui.adsabs.harvard.edu/abs/2013ApJ...763L..15M} {763, L15}

\bibitem[\protect\citeauthoryear{{Morscher}, {Pattabiraman}, {Rodriguez},
  {Rasio}  \& {Umbreit}}{{Morscher} et~al.}{2015}]{Morscher2015}
{Morscher} M.,  {Pattabiraman} B.,  {Rodriguez} C.,  {Rasio} F.~A.,   {Umbreit}
  S.,  2015, \mn@doi [\apj] {10.1088/0004-637X/800/1/9}, 800, 9

\bibitem[\protect\citeauthoryear{{Noyola}, {Gebhardt}  \& {Bergmann}}{{Noyola}
  et~al.}{2008}]{Noyola2008}
{Noyola} E.,  {Gebhardt} K.,   {Bergmann} M.,  2008, \mn@doi [\apj]
  {10.1086/529002}, 676, 1008

\bibitem[\protect\citeauthoryear{{Paduano} et~al.,}{{Paduano}
  et~al.}{2024}]{Paduano2024}
{Paduano} A.,  et~al., 2024, \mn@doi [\apj] {10.3847/1538-4357/ad0e68}, 961, 54

\bibitem[\protect\citeauthoryear{{Perera} et~al.,}{{Perera}
  et~al.}{2017}]{Perera2017a}
{Perera} B.~B.~P.,  et~al., 2017, \mn@doi [\mnras] {10.1093/mnras/stx501}, 468,
  2114

\bibitem[\protect\citeauthoryear{{Peuten}, {Zocchi}, {Gieles}, {Gualandris}  \&
  {H{\'e}nault-Brunet}}{{Peuten} et~al.}{2016}]{Peuten2016}
{Peuten} M.,  {Zocchi} A.,  {Gieles} M.,  {Gualandris} A.,
  {H{\'e}nault-Brunet} V.,  2016, \mn@doi [\mnras] {10.1093/mnras/stw1726},
  462, 2333

\bibitem[\protect\citeauthoryear{{Peuten}, {Zocchi}, {Gieles}  \&
  {H{\'e}nault-Brunet}}{{Peuten} et~al.}{2017}]{Peuten2017}
{Peuten} M.,  {Zocchi} A.,  {Gieles} M.,   {H{\'e}nault-Brunet} V.,  2017,
  \mn@doi [\mnras] {10.1093/mnras/stx1311}, 470, 2736

\bibitem[\protect\citeauthoryear{{Portegies Zwart} \& {McMillan}}{{Portegies
  Zwart} \& {McMillan}}{2000}]{PortegiesZwart2000}
{Portegies Zwart} S.~F.,  {McMillan} S.~L.~W.,  2000, \mn@doi [\apjl]
  {10.1086/312422}, 528, L17

\bibitem[\protect\citeauthoryear{{Rodriguez}, {Morscher}, {Wang}, {Chatterjee},
  {Rasio}  \& {Spurzem}}{{Rodriguez} et~al.}{2016}]{Rodriguez2016}
{Rodriguez} C.~L.,  {Morscher} M.,  {Wang} L.,  {Chatterjee} S.,  {Rasio}
  F.~A.,   {Spurzem} R.,  2016, \mn@doi [\mnras] {10.1093/mnras/stw2121}, 463,
  2109

\bibitem[\protect\citeauthoryear{{Rodriguez} et~al.,}{{Rodriguez}
  et~al.}{2022}]{Rodriguez2022}
{Rodriguez} C.~L.,  et~al., 2022, \mn@doi [\apjs] {10.3847/1538-4365/ac2edf},
  258, 22

\bibitem[\protect\citeauthoryear{Rui, Kremer, Weatherford, Chatterjee, Rasio,
  Rodriguez  \& Ye}{Rui et~al.}{2021a}]{cmctoolkit}
Rui N.~Z.,  Kremer K.,  Weatherford N.~C.,  Chatterjee S.,  Rasio F.~A.,
  Rodriguez C.~L.,   Ye C.~S.,  2021a, NicholasRui/cmctoolkit: First release,
  \mn@doi{10.5281/zenodo.4579951}, \url
  {https://doi.org/10.5281/zenodo.4579951}

\bibitem[\protect\citeauthoryear{{Rui}, {Weatherford}, {Kremer}, {Chatterjee},
  {Fragione}, {Rasio}, {Rodriguez}  \& {Ye}}{{Rui} et~al.}{2021b}]{Rui2021b}
{Rui} N.~Z.,  {Weatherford} N.~C.,  {Kremer} K.,  {Chatterjee} S.,  {Fragione}
  G.,  {Rasio} F.~A.,  {Rodriguez} C.~L.,   {Ye} C.~S.,  2021b, \mn@doi
  [Research Notes of the American Astronomical Society]
  {10.3847/2515-5172/abee77}, 5, 47

\bibitem[\protect\citeauthoryear{{Rui}, {Kremer}, {Weatherford}, {Chatterjee},
  {Rasio}, {Rodriguez}  \& {Ye}}{{Rui} et~al.}{2021c}]{Rui2021a}
{Rui} N.~Z.,  {Kremer} K.,  {Weatherford} N.~C.,  {Chatterjee} S.,  {Rasio}
  F.~A.,  {Rodriguez} C.~L.,   {Ye} C.~S.,  2021c, \mn@doi [\apj]
  {10.3847/1538-4357/abed49}, 912, 102

\bibitem[\protect\citeauthoryear{{Sigurdsson} \& {Hernquist}}{{Sigurdsson} \&
  {Hernquist}}{1993}]{Sigurdsson1993}
{Sigurdsson} S.,  {Hernquist} L.,  1993, \mn@doi [\nat] {10.1038/364423a0},
  364, 423

\bibitem[\protect\citeauthoryear{{Sollima} \& {Baumgardt}}{{Sollima} \&
  {Baumgardt}}{2017}]{Sollima2017}
{Sollima} A.,  {Baumgardt} H.,  2017, \mn@doi [\mnras] {10.1093/mnras/stx1856},
  471, 3668

\bibitem[\protect\citeauthoryear{{Sollima}, {Bellazzini}  \& {Lee}}{{Sollima}
  et~al.}{2012}]{Sollima2012}
{Sollima} A.,  {Bellazzini} M.,   {Lee} J.~W.,  2012, \mn@doi [\apj]
  {10.1088/0004-637X/755/2/156}, 755, 156

\bibitem[\protect\citeauthoryear{{Sollima} et~al.,}{{Sollima}
  et~al.}{2016}]{Sollima2016}
{Sollima} A.,  et~al., 2016, \mn@doi [\mnras] {10.1093/mnras/stw1779}, 462,
  1937

\bibitem[\protect\citeauthoryear{{Speagle}}{{Speagle}}{2020}]{Speagle2020}
{Speagle} J.~S.,  2020, \mn@doi [\mnras] {10.1093/mnras/staa278}, 493, 3132

\bibitem[\protect\citeauthoryear{{Spera}, {Mapelli}  \& {Bressan}}{{Spera}
  et~al.}{2015}]{Spera2015}
{Spera} M.,  {Mapelli} M.,   {Bressan} A.,  2015, \mn@doi [\mnras]
  {10.1093/mnras/stv1161}, 451, 4086

\bibitem[\protect\citeauthoryear{{Spitzer}}{{Spitzer}}{1969}]{Spitzer1969}
{Spitzer} Lyman J.,  1969, \mn@doi [\apjl] {10.1086/180451}, 158, L139

\bibitem[\protect\citeauthoryear{Spitzer}{Spitzer}{1987}]{Spitzer1987}
Spitzer L.~S.,  1987, Dynamical Evolution of Globular Clusters.
Princeton University Press

\bibitem[\protect\citeauthoryear{{Strader}, {Chomiuk}, {Maccarone},
  {Miller-Jones}  \& {Seth}}{{Strader} et~al.}{2012}]{Strader2012}
{Strader} J.,  {Chomiuk} L.,  {Maccarone} T.~J.,  {Miller-Jones} J. C.~A.,
  {Seth} A.~C.,  2012, \mn@doi [\nat] {10.1038/nature11490}, 490, 71

\bibitem[\protect\citeauthoryear{{Torniamenti}, {Gieles}, {Penoyre},
  {Jerabkova}, {Wang}  \& {Anders}}{{Torniamenti}
  et~al.}{2023}]{Torniamenti2023}
{Torniamenti} S.,  {Gieles} M.,  {Penoyre} Z.,  {Jerabkova} T.,  {Wang} L.,
  {Anders} F.,  2023, \mn@doi [\mnras] {10.1093/mnras/stad1925}, 524, 1965

\bibitem[\protect\citeauthoryear{{Trager}, {King}  \& {Djorgovski}}{{Trager}
  et~al.}{1995}]{Trager1995}
{Trager} S.~C.,  {King} I.~R.,   {Djorgovski} S.,  1995, \mn@doi [\aj]
  {10.1086/117268}, 109, 218

\bibitem[\protect\citeauthoryear{{Vasiliev} \& {Baumgardt}}{{Vasiliev} \&
  {Baumgardt}}{2021}]{Vasiliev2021}
{Vasiliev} E.,  {Baumgardt} H.,  2021, \mn@doi [\mnras]
  {10.1093/mnras/stab1475}, 505, 5978

\bibitem[\protect\citeauthoryear{{Virtanen} et~al.,}{{Virtanen}
  et~al.}{2020}]{Virtanen2020}
{Virtanen} P.,  et~al., 2020, \mn@doi [Nature Methods]
  {10.1038/s41592-019-0686-2}, \href
  {https://ui.adsabs.harvard.edu/abs/2020NatMe..17..261V} {17, 261}

\bibitem[\protect\citeauthoryear{{Vitral} \& {Mamon}}{{Vitral} \&
  {Mamon}}{2021}]{Vitral2021}
{Vitral} E.,  {Mamon} G.~A.,  2021, \mn@doi [\aap]
  {10.1051/0004-6361/202039650}, 646, A63

\bibitem[\protect\citeauthoryear{{Vitral}, {Kremer}, {Libralato}, {Mamon}  \&
  {Bellini}}{{Vitral} et~al.}{2022}]{Vitral2022}
{Vitral} E.,  {Kremer} K.,  {Libralato} M.,  {Mamon} G.~A.,   {Bellini} A.,
  2022, \mn@doi [\mnras] {10.1093/mnras/stac1337}, 514, 806

\bibitem[\protect\citeauthoryear{{Vitral}, {Libralato}, {Kremer}, {Mamon},
  {Bellini}, {Bedin}  \& {Anderson}}{{Vitral} et~al.}{2023}]{Vitral2023}
{Vitral} E.,  {Libralato} M.,  {Kremer} K.,  {Mamon} G.~A.,  {Bellini} A.,
  {Bedin} L.~R.,   {Anderson} J.,  2023, \mn@doi [\mnras]
  {10.1093/mnras/stad1068}, 522, 5740

\bibitem[\protect\citeauthoryear{{Watkins}, {van der Marel}, {Bellini}  \&
  {Anderson}}{{Watkins} et~al.}{2015}]{Watkins2015}
{Watkins} L.~L.,  {van der Marel} R.~P.,  {Bellini} A.,   {Anderson} J.,  2015,
  \mn@doi [\apj] {10.1088/0004-637X/803/1/29}, 803, 29

\bibitem[\protect\citeauthoryear{{Weatherford}, {Chatterjee}, {Rodriguez}  \&
  {Rasio}}{{Weatherford} et~al.}{2018}]{Weatherford2018}
{Weatherford} N.~C.,  {Chatterjee} S.,  {Rodriguez} C.~L.,   {Rasio} F.~A.,
  2018, \mn@doi [\apj] {10.3847/1538-4357/aad63d}, 864, 13

\bibitem[\protect\citeauthoryear{{Weatherford}, {Chatterjee}, {Kremer}  \&
  {Rasio}}{{Weatherford} et~al.}{2020}]{Weatherford2020}
{Weatherford} N.~C.,  {Chatterjee} S.,  {Kremer} K.,   {Rasio} F.~A.,  2020,
  \mn@doi [\apj] {10.3847/1538-4357/ab9f98}, 898, 162

\bibitem[\protect\citeauthoryear{{Ye}, {Kremer}, {Rodriguez}, {Rui},
  {Weatherford}, {Chatterjee}, {Rasio}  \& {Fragione}}{{Ye}
  et~al.}{2022}]{Ye2022}
{Ye} C.~S.,  {Kremer} K.,  {Rodriguez} C.~L.,  {Rui} N.~Z.,  {Weatherford}
  N.~C.,  {Chatterjee} S.,  {Rasio} F.~A.,   {Fragione} G.,  2022, \mn@doi
  [\apj] {10.3847/1538-4357/ac5b0b}, 931, 84

\bibitem[\protect\citeauthoryear{{Zocchi}, {Gieles}  \&
  {H{\'e}nault-Brunet}}{{Zocchi} et~al.}{2017}]{Zocchi2017}
{Zocchi} A.,  {Gieles} M.,   {H{\'e}nault-Brunet} V.,  2017, \mn@doi [\mnras]
  {10.1093/mnras/stx316}, 468, 4429

\bibitem[\protect\citeauthoryear{{Zocchi}, {Gieles}  \&
  {H{\'e}nault-Brunet}}{{Zocchi} et~al.}{2019}]{Zocchi2019}
{Zocchi} A.,  {Gieles} M.,   {H{\'e}nault-Brunet} V.,  2019, \mn@doi [\mnras]
  {10.1093/mnras/sty1508}, 482, 4713

\bibitem[\protect\citeauthoryear{{de Boer}, {Gieles}, {Balbinot},
  {H{\'e}nault-Brunet}, {Sollima}, {Watkins}  \& {Claydon}}{{de Boer}
  et~al.}{2019}]{deBoer2019}
{de Boer} T.~J.~L.,  {Gieles} M.,  {Balbinot} E.,  {H{\'e}nault-Brunet} V.,
  {Sollima} A.,  {Watkins} L.~L.,   {Claydon} I.,  2019, \mn@doi [\mnras]
  {10.1093/mnras/stz651}, 485, 4906

\bibitem[\protect\citeauthoryear{{van der Marel} \& {Anderson}}{{van der Marel}
  \& {Anderson}}{2010}]{vanderMarel2010}
{van der Marel} R.~P.,  {Anderson} J.,  2010, \mn@doi [\apj]
  {10.1088/0004-637X/710/2/1063}, 710, 1063

\makeatother
\end{thebibliography}

%---------------------------------------------------------------------------
% APPENDICES
%---------------------------------------------------------------------------

\appendix

\section{Results of the validation of BH population inference}

\begin{table*}
    \renewcommand*{\arraystretch}{1.4}
    \centering
    \footnotesize
    \begin{tabular}{lllll}
    \hline
        \multicolumn{1}{c}{\multirow{2}{*}{CMC Model}} & \multicolumn{1}{c}{\multirow{2}{*}{Snapshot}} & \multirow{2}{*}{\shortstack[c]{Closest Milky \\ Way Analogue}} & \multicolumn{1}{c}{\multirow{2}{*}{\(f_{\mathrm{BH, true}}\ \left[\%\right]\)}} & \multicolumn{1}{c}{\multirow{2}{*}{\(\fbh\ \left[\%\right]\)}} \\
        & & & & \\
    \hline

        N4e5-rv1-rg2-Z0.0002                           & 248                                           & \NGC6558                                     & 0.0   & $0.000\substack{+0.003 \\ -0.000}$ \\
        N4e5-rv1-rg8-Z0.0002                           & 217                                           & \NGC5946                                     & 0.008 & $0\substack{+0         \\ -0}$             \\
        N4e5-rv1-rg8-Z0.0002                           & 217                                           & \NGC5986                                     & 0.008 & $0.03\substack{+0.02   \\ -0.02}$    \\
        N4e5-rv1-rg8-Z0.0002                           & 252                                           & \NGC6544                                     & 0.015 & $0.004\substack{+0.007 \\ -0.004}$ \\
        N4e5-rv2-rg2-Z0.002                            & 165                                           & \NGC6352                                     & 0.152 & $0.15\substack{+0.08   \\ -0.07}$    \\
        N4e5-rv2-rg8-Z0.0002\(^\ast\)                  & 160                                           & \NGC6584                                     & 0.222 & $0.13\substack{+0.06   \\ -0.05}$    \\
        N4e5-rv2-rg8-Z0.0002                           & 165                                           & \NGC6981                                     & 0.147 & $0.13\substack{+0.08   \\ -0.07}$    \\
        N4e5-rv4-rg8-Z0.0002\(^\ast\)                  & 128                                           & \NGC5897                                     & 0.786 & $0.3\substack{+0.2     \\ -0.2}$       \\
        N4e5-rv4-rg8-Z0.0002                           & 134                                           & \NGC288                                      & 0.624 & $0.6\substack{+0.3     \\ -0.3}$       \\
        N8e5-rv0.5-rg2-Z0.0002                         & 608                                           & \NGC6355                                     & 0.0   & $0.019\substack{+0.011 \\ -0.008}$ \\
        N8e5-rv0.5-rg2-Z0.0002                         & 608                                           & \NGC6681                                     & 0.0   & $0.02\substack{+0.01   \\ -0.01}$    \\
        N8e5-rv0.5-rg2-Z0.002                          & 516                                           & \NGC6624                                     & 0.0   & $0\substack{+0         \\ -0}$             \\
        N8e5-rv1-rg2-Z0.0002                           & 375                                           & \NGC6293                                     & 0.064 & $0.10\substack{+0.05   \\ -0.03}$    \\
        N8e5-rv1-rg2-Z0.0002                           & 375                                           & \NGC6453                                     & 0.064 & $0.07\substack{+0.01   \\ -0.02}$    \\
        N8e5-rv1-rg2-Z0.0002                           & 375                                           & \NGC6522                                     & 0.064 & $0.24\substack{+0.06   \\ -0.06}$    \\
        N8e5-rv1-rg2-Z0.002                            & 340                                           & \NGC6316                                     & 0.229 & $0.09\substack{+0.02   \\ -0.03}$    \\
        N8e5-rv1-rg2-Z0.002                            & 340                                           & \NGC6539                                     & 0.229 & $0.13\substack{+0.03   \\ -0.03}$    \\
        N8e5-rv1-rg2-Z0.002                            & 345                                           & \NGC6569                                     & 0.18  & $0.09\substack{+0.03   \\ -0.04}$    \\
        N8e5-rv1-rg2-Z0.002                            & 345                                           & \NGC6637                                     & 0.18  & $0.10\substack{+0.03   \\ -0.03}$    \\
        N8e5-rv1-rg2-Z0.002                            & 345                                           & \NGC6642                                     & 0.18  & $0.25\substack{+0.10   \\ -0.09}$    \\
        N8e5-rv1-rg2-Z0.002                            & 354                                           & \NGC6638                                     & 0.179 & $0.09\substack{+0.02   \\ -0.03}$    \\
        N8e5-rv1-rg2-Z0.002                            & 358                                           & \NGC6401                                     & 0.175 & $0.09\substack{+0.02   \\ -0.02}$    \\
        N8e5-rv1-rg2-Z0.002                            & 362                                           & \NGC6256                                     & 0.155 & $0.09\substack{+0.03   \\ -0.02}$    \\
        N8e5-rv1-rg2-Z0.002                            & 362                                           & \NGC6304                                     & 0.155 & $0.09\substack{+0.02   \\ -0.02}$    \\
        N8e5-rv1-rg20-Z0.0002                          & 394                                           & \NGC1904                                     & 0.154 & $0.08\substack{+0.02   \\ -0.04}$    \\
        N8e5-rv1-rg8-Z0.0002                           & 379                                           & \NGC6284                                     & 0.188 & $0.09\substack{+0.03   \\ -0.04}$    \\
        N8e5-rv1-rg8-Z0.0002                           & 386                                           & \NGC6779                                     & 0.153 & $0.08\substack{+0.03   \\ -0.04}$    \\
        N8e5-rv1-rg8-Z0.0002                           & 386                                           & \NGC6934                                     & 0.153 & $0.09\substack{+0.03   \\ -0.04}$    \\
        N8e5-rv1-rg8-Z0.002\(^\ast\)                   & 366                                           & \NGC6760                                     & 0.234 & $0.18\substack{+0.05   \\ -0.04}$    \\
        N8e5-rv2-rg2-Z0.0002                           & 315                                           & \NGC6218                                     & 0.456 & $0.3\substack{+0.1     \\ -0.1}$       \\
        N8e5-rv2-rg2-Z0.002\(^\ast\)                   & 273                                           & \NGC6723                                     & 0.977 & $0.3\substack{+0.1     \\ -0.1}$       \\
        N8e5-rv2-rg2-Z0.002\(^\ast\)                   & 286                                           & \NGC6712                                     & 0.798 & $0.17\substack{+0.08   \\ -0.10}$    \\
        N8e5-rv2-rg20-Z0.0002\(^\ast\)                 & 296                                           & \NGC1261                                     & 0.735 & $0.09\substack{+0.07   \\ -0.07}$    \\
        N8e5-rv2-rg8-Z0.0002                           & 299                                           & \NGC3201                                     & 0.694 & $0.5\substack{+0.1     \\ -0.1}$       \\
        N8e5-rv4-rg8-Z0.0002\(^\ast\)                  & 194                                           & \NGC4372                                     & 2.218 & $0.3\substack{+0.2     \\ -0.2}$       \\
        N8e5-rv4-rg8-Z0.0002\(^\ast\)                  & 194                                           & \NGC6101                                     & 2.218 & $0.4\substack{+0.3     \\ -0.3}$       \\
        N1.6e6-rv0.5-rg2-Z0.002                        & 875                                           & \NGC6440                                     & 0.133 & $0.32\substack{+0.04   \\ -0.05}$    \\
        N1.6e6-rv1-rg2-Z0.0002                         & 694                                           & \NGC6402                                     & 0.898 & $0.8\substack{+0.2     \\ -0.2}$       \\
        N1.6e6-rv1-rg2-Z0.0002                         & 727                                           & \NGC6656                                     & 0.734 & $1.3\substack{+0.2     \\ -0.2}$       \\
        N1.6e6-rv1-rg2-Z0.0002                         & 737                                           & \NGC6541                                     & 0.679 & $0.8\substack{+0.1     \\ -0.1}$       \\
        N1.6e6-rv1-rg2-Z0.0002                         & 755                                           & \NGC6333                                     & 0.584 & $0.9\substack{+0.1     \\ -0.1}$       \\
        N1.6e6-rv1-rg2-Z0.0002                         & 755                                           & \NGC6342                                     & 0.584 & $1.7\substack{+0.4     \\ -0.3}$       \\
        N1.6e6-rv1-rg8-Z0.002\(^\ast\)                 & 698                                           & \NGC6356                                     & 0.715 & $0.10\substack{+0.06   \\ -0.08}$    \\
    
    \hline
    \end{tabular}
    \caption{Results of the validation fits described in \Cref{sec:validation}.
             The ``True'' model values (\(f_{\mathrm{BH, true}}\)) and median
             and \(1\sigma\) credibility intervals inferred by the fits to
             mock observations (\(f_{\mathrm{BH, infer}}\)) are shown.
             The model names (as given in the CMC cluster catalogue) indicate
             the initial number of particles, virial radii, galactocentric
             radii and metallicity of each CMC model.
             The Milky Way cluster to which each snapshot was matched is also
             given.
             All dynamically young snapshots which fall into the
             suspect region identified in \Cref{sec:validation} are denoted by
             an asterisk.
             }
    \label{table:val_results}
\end{table*}
%---------------------------------------------------------------------------
% Don't change these lines
%---------------------------------------------------------------------------

\bsp    % typesetting comment
\label{lastpage}
\end{document}